\documentclass[journal]{IEEEtran}
\usepackage{booktabs}
\usepackage{tabularx}
\usepackage{amsmath,amssymb}
\usepackage{graphicx}
\usepackage{epstopdf}
\usepackage{float}
\usepackage{multirow}
\usepackage{cite}
\usepackage{booktabs}
\usepackage{subfigure}
\usepackage{xcolor}
\usepackage{amsfonts}
\usepackage{mathrsfs}
\usepackage{pifont}
\usepackage{stfloats}
\usepackage{color}
\usepackage{booktabs}
\usepackage{caption}
\usepackage{algorithm}
\usepackage{caption}
\usepackage{algorithmicx}
\usepackage{algpseudocode}
\renewcommand{\t}{^{\mbox{\tiny {T}}}}

\IEEEoverridecommandlockouts                              
\newcommand{\eproof}{\hfill\rule{2mm}{2mm}}

\newcommand{\bstate}{\begin{state} }
	\newcommand{\estate}{ \hfill  \rule{1mm}{2mm}\end{state}}

\newcommand{\bass}{\begin{ass} }
	\newcommand{\eass}{ \hfill  \rule{1mm}{2mm}\end{ass}}

\newcommand{\brem}{ \begin{remark}  }
	\newcommand{\erem}{\hfill \rule{1mm}{2mm}
\end{remark} }
\newcommand{\bthm}{\begin{theorem}  }
	\newcommand{\ethm}{ \hfill  \rule{1mm}{2mm}
\end{theorem} }
\newcommand{\blem}{\begin{lemma}  }
	\newcommand{\elem}{ \hfill \rule{1mm}{2mm}
\end{lemma} }
\newcommand{\bcorollary}{\begin{corollary}  }
	\newcommand{\ecorollary}{  \hfill \rule{1mm}{2mm}
\end{corollary} }
\newcommand{\bdefn}{\begin{definition}}
	\newcommand{\edefn}{  \hfill \rule{1mm}{2mm}
\end{definition} }
\newcommand{\bproposition}{\begin{proposition} }
	\newcommand{\eproposition}{\hfill \rule{1mm}{2mm}
\end{proposition} }
\newcommand{\bexample}{\begin{example} \rm}
	\newcommand{\eexample}{ \hfill \rule{1mm}{2mm}
\end{example} }

\newcommand{\bcon}{\begin{condition} \rm}
	\newcommand{\econ}{ \hfill \rule{1mm}{2mm}
\end{condition} }

\newcommand{\proofnow}{\noindent{\bf Proof: }}
\newcommand{\prooflater}[1]{\noindent{\bf Proof of #1: }}

\newtheorem{theorem}{\bf Theorem}[section]
\newtheorem{ass}{\bf Assumption}[section]
\newtheorem{lemma}{\bf Lemma}[section]
\newtheorem{definition}{\bf Definition}[section]
\newtheorem{remark}{\bf Remark}[section]
\newtheorem{corollary}{\bf Corollary}[section]
\newtheorem{proposition}{\bf Proposition}[section]
\newtheorem{example}{\bf Example}[section]
\newtheorem{condition}{\bf Condition}[section]
\newtheorem{state}{\bf Assumption}[section]

\newcommand{\blue}{\color{black}}

\usepackage{fancyhdr}
\begin{document}
\title{Adaptive Event-triggered Control with Sampled Transmitted Output and Controller Dynamics}
\author{Gewei Zuo, Lijun Zhu
\thanks{The authors are with School of Artificial Intelligence and Automation, Huazhong University of Science and Technology, Wuhan 430072, China. L. Zhu is also with Key Laboratory of Imaging Processing and Intelligence Control (e-mail:gwzuo@hust.edu.cn; ljzhu@hust.edu.cn). 
}}

\markboth{}{ \MakeLowercase{\textit{et al.}}: }

\maketitle

\begin{abstract}
The event-triggered control with intermittent output can reduce the communication burden between the controller and plant side over the network.   It has been exploited for adaptive output feedback control of uncertain nonlinear systems in the literature,
however the controller  must partially reside at the plant side  where  the computation capacity is required.  In this paper,   all controller components are moved to the controller side and
their dynamics use sampled states rather than continuous one  with the benefit of directly estimating next  triggering instance of some conditions and avoiding constantly checking event condition at the controller side. However, these  bring two major challenges. First, the virtual input designed in the dynamic filtering technique for the stabilization is no longer differentiable.
Second, the  plant output is sampled to transmit at plant side and sampled again at controller side to construct the controller, and the two asynchronous samplings  make  the analysis more involving.
This paper solves these two   issues   by  introducing a new state observer to simplify the adaptive law, a set of continuous companion variables for stability analysis   and a new lemma quantifying the error bound between actual output signal and sampled transmitted output.
 It is theoretically guaranteed  that all internal signals in the closed-loop system are  semiglobally  bounded and
 the output is practically stabilized to the origin. Finally, the numerical
  simulation   illustrates the effectiveness of proposed scheme.
\end{abstract}
\begin{IEEEkeywords}
Event-triggered control; Sampled transmitted output; Sampled controller  dynamics;   Dynamic filtering.
\end{IEEEkeywords}

\IEEEpeerreviewmaketitle

\section{Introduction}
Network control, as an integrated technology of computation, communication and physical process, plays an increasingly important role in critical infrastructure, 
due to its advantages in improving control efficiency and reducing control cost \cite{lian2001performance24}. To achieve more efficient network control, the key lies in how to reduce the communication and computational burdens on the premise of ensuring the closed-loop system's stability. Event-triggered control has been proved effective in reducing these burdens and  used to  control  satellites \cite{liu2019neural16}, unmanned aerial vehicles \cite{song2019event14} and robots \cite{yang2020event15}. Compared with traditional sample-date control, the event-triggered control can effectively reduce signal transmission from controller side to plant side \cite{aastrom1999comparison12}.

A lot of progress has been made for  the
event-triggered control,   as pointed out in \cite{liu2019neural16,song2019event14, yang2020event15,aastrom1999comparison12}, \cite{huang2019adaptive01,tabuada2007event13,zhu2019new10, xing2016event04,xing2018event02,zhu2021event11,li2020adaptive05,zhu2022sampling23, zhang2021adaptive03} and references therein, but there are still some limitations. In \cite{tabuada2007event13,zhu2019new10}, the event-triggered control algorithms are proposed for a class of  systems with and without uncertainties. In \cite{xing2016event04}, adaptive event-triggered control for strict-feedback systems is considered. These schemes are based on real-time monitoring of system state, that means although control signal is not transmitted to plant side during the triggering interval, event detector  located at controller side must continuously access  state signals at the plant side to
 determine whether the control signal needs to be updated.
  This undoubtedly increases the communication in the channel from the plant side to  controller side, and  also consumes additional network bandwidth. Although output feedback based event-triggered control algorithms are proposed in \cite{xing2018event02,zhu2021event11,li2020adaptive05,zhu2022sampling23}, the real-time monitoring behavior still exists.
  This problem has been recently addressed in \cite{zhang2021adaptive03} using dynamic filtering technique, where an event detector is constructed at the plant side to determine whether the output signal needs to be transmitted. However, the dynamic filter as  the controller component must  reside at the plant side such that the dynamic filtering technique can be applicable. Therefore, the computation capacity is also needed at the plant side.  Based on \cite{zhang2021adaptive03}, intermittent feedback control algorithm is further extended to interconnected systems \cite{zhang2022decentralized30,sun2023decentralized29}, multi-agent systems \cite{sun2022distributed28}, while the constant checking behavior of trigger condition at plant side still exists.

This paper proposes an event-triggered adaptive output feedback control algorithm.
Compared with \cite{zhang2021adaptive03}, we move all components of the  controller to the controller side and thus remove the requirement of computation capacity at the plant side. However, it brings a challenge that the virtual input designed in the dynamic filtering technique is no longer 
differentiable.
The main features and contributions of this paper are summarized as follows. (1) Two event detectors are constructed. One is located at the plant side to trigger output transmission, and the other one resides at the controller side to trigger the update of  control signal and  the controller dynamics. As a result, the transmission between controller side and plant side is intermittent, and the corresponding continuous variables can be calculated by simple algebraic equation rather than numerical integration of the vector function. (2) The controller  uses the  value of  transmitted output  from  the plant side  sampled at the time instance triggered by the event detector  at the controller side. A new lemma is introduced to analyze the error bound between actual output signal   and   sampled transmitted output, which plays an important role in  stability analysis. Compared to \cite{zhang2021adaptive03,zhang2022decentralized30,sun2023decentralized29,sun2022distributed28}, we only need to check whether there is a new output signal arrival, and calculate the variation of new output signal to determine the next triggering instance, thus the constant checking behavior of event condition at controller side is avoided. (3) A set of companion variables is introduced and backstepping method with dynamic filter is adopted
to solve the issue that the virtual input is no longer differentiable caused by sampled output and controller dynamics. Semi-globally practical stability is guaranteed, which is a sacrifice of control accuracy for the introduction of two event detectors when compared to asymptotic stability. (4) A new state estimator and  a novel coordinate transformation for the backstepping design procedure such that  parameter estimation law is only  introduced  in the first step, thus simplifying the adaptive law structure compared with \cite{zhang2021adaptive03}.

The rest of paper is organized as follows. Section \ref{sec:PF} presents the system model, event-triggered mechanism and control objective. Section \ref{sec:MR} proposes the triggering conditions, elaborates the state observer, adaptive law and dynamic filter, and proves the closed-loop system's stability. Section \ref{sec:sim} verifies the effectiveness of proposed control algorithm with a numerical example, and Section \ref{sec:con} concludes the paper.






{\blue \section{Notations and Problem Formulation \label{sec:PF}}}
{\blue \subsection{Notations}
$\mathbb R$ and $\mathbb{R}^n$ denote the set of real numbers and the $n$-dimensional Euclidean space, respectively. $\mathbb{R}^{n\times n}$ denotes the $n$-dimensional square matrix whose elements are all real numbers. $\mathbb N$ denotes the set of nonnegative integers.  Unless otherwise specified, $\|A\|$ denotes the $2$-norm of a vector or matrix. $\mathcal C^1$ is the set of all first-order differentiable functions. }
\subsection{Problem Formulation}
Consider the following nonlinear system in output-feedback form, as studied in \cite{zhang2021adaptive03,xing2018event02}, and Section 7 of \cite{krstic1995nonlinear06},
\begin{equation}\label{eq1}
\begin{aligned}
\dot{x}_i&=x_{i+1}+\theta\psi_i(y),\ i=1,\cdots,n-1\\
\dot{x}_n&=u+\theta\psi_n(y)\\
y&=x_1
\end{aligned}
\end{equation}
where {\blue $x=[x_1,\cdots,x_n]\t\in\mathbb{R}^n$}, $y\in\mathbb{R}$, $u\in\mathbb{R}$ denote state, output and control input, respectively. $\theta\in\mathbb{R}$ is an unknown constant 
satisfying
$\vert \theta\vert \leq \bar \theta$. $\psi_i(y),i=1,\cdots,n$ are known nonlinear functions with $\psi_i(0)$ not necessarily being zero
and satisfy the  Lipschitz condition.

\bass\label{ass1}
The function $\psi_i(y)$ is $\mathcal{C}^1$ function and  satisfies the  Lipschitz continuity, i.e., there exist  known parameters $L_i$ such that
\begin{equation}\label{eq2}
\vert\psi_i(y)-\psi_i(y^*) \vert\leq  L_i\vert y-y^*\vert,\; \forall y ,y^*\in \mathbb{D}\subset \mathbb{R}
\end{equation}
for $i=1,\cdots,n$ where $\mathbb{D}$ is a closed set.
And the first partial derivative with respect to $y$ satisfies
\begin{equation}\label{eq136}
\vert {\partial \psi_i(y)}/{\partial y}\vert \leq \Psi_i,\;\forall y \in \mathbb{D}\subset \mathbb{R}
\end{equation}
where $\Psi_i>0$.
\eass
{\blue \brem
In this paper, our proposed algorithm is also applicable for the case that different relative degrees contain different unknown scalar parameters, i.e,
 $\dot x_i=x_{i+1}+\theta_i\psi_i(y),i=1,\cdots,n-1$, $\dot{x}_n = u+\theta_n \psi_n(y)$, $y=x_1$
 with $\theta_i,\psi_i(y)\in \mathbb R$ for $i=1,\cdots,n$ or even they contain  different unknown vector parameters, i.e, $\dot x_i=x_{i+1}+\theta_i\t\psi_i(y)$, $\dot x_n =u+\theta_n\t \psi_n(y)$, $y=x_1$ with $\theta_i,\psi_i(y)\in\mathbb R^m$.
  For the simplicity of presentation, we only consider the system has one unknown scalar parameter $\theta$. The proposed algorithm is also effective in output-feedback chain-integrator system, i.e, $\dot x_i = x_{i+1},i=1,\cdots,n-1$, $\dot x_n = u+\theta_n \t\psi_n(y)$, $y=x_1$, which is a special case of (\ref{eq1}) with  $\psi_i (y)=0 $ for $i=1,\cdots,n-1$.
It's worth noting that the Lipschitz continuity of $\psi_i(y)$ is also required for event-triggered control system   in \cite{huang2019adaptive01,zhang2021adaptive03}.
\erem }

{\blue For continuous-time stabilization of the system (\ref{eq1}), to tackle the output measurement and the existence of uncertain parameter,  the  state observer,  adaptation dynamics and dynamic filter are usually proposed, respectively, as
\begin{equation}
\dot {\hat{x}}=\chi(\hat{x},y),\quad
\dot {\hat{\theta}}=\vartheta(\hat\theta,\hat{x},y),\quad \dot { \alpha}_f=\varsigma(\alpha_f,\hat\theta,\hat{x},y).\label{eq:xt}
\end{equation}
 Together with (\ref{eq:xt}), the   controller can be constructed as
\begin{equation}
u=\kappa( \alpha_f,\hat\theta,\hat{x},y). \label{eq:u}
\end{equation}
The $\alpha_f$-dynamics can be designed recursively by the backstepping technique \cite{krstic1992adaptive27,swaroop2000dynamic09}.}

\begin{figure}[!htp]
	\centering
	\includegraphics[width=1\linewidth]{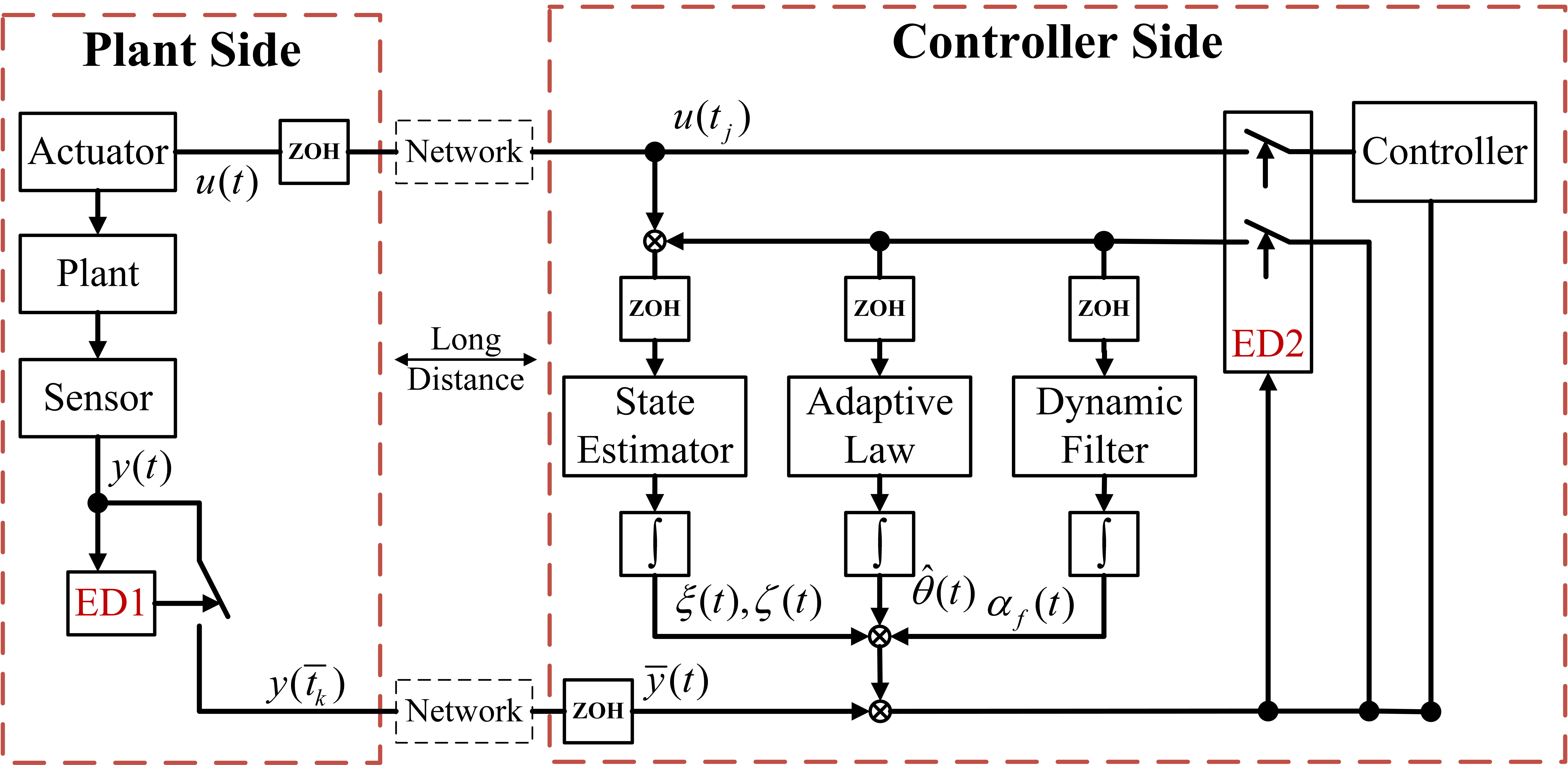}
\captionsetup{font={small}}
\vspace{2pt}
	\caption{\small{Illustration of the proposed  output feedback control strategy with output, controller, state observer, adaptive law, dynamic filter, and  event detectors. $\otimes$ denotes the combination of different signals; $\bullet$ denotes the replication of signal. ZOH is zero-order holder which can transform discrete signals into piecewise-continuous signals.}}
	\label{fig1}
\end{figure}

In this paper, we consider the stabilization control in a networked environment, that is the plant (\ref{eq1}) and controller composed of (\ref{eq:xt}) and (\ref{eq:u})  reside in the two different locations and communicate over the network, as illustrated in Fig \ref{fig1}. The output $y$, controller $u$ are transmitted over the network.
In order to reduce the
communication,  information  transmission  is  schedule by the event
detectors  at the plant  (ED1) and controller sides (ED2).
Denote the triggering sequence of event detectors at the plant and controller side as $\{\bar t_1,\cdots,\bar t_k\}$ and $\{t_1,\cdots,t_j\}$, respectively, where $k,j\in\mathbb{N}$. Then, the information, that is received by the controller and plant sides and both kept by corresponding zero-order holder (ZOH),  is denoted respectively as
\begin{equation}\label{eq:bar_y_and_u}
\begin{aligned}
\bar{y}(t)&=y(\bar t_{k}),\  t\in [\bar t_{k},\bar t_{k+1})\\
u(t)&=\kappa( \alpha_f(t_j),\hat\theta(t_j),\hat{x}(t_j),\bar y(t_j)),\  t\in [ t_{j}, t_{j+1})
\end{aligned}
\end{equation}
where $\bar y(t_j)$ is the value of transmitted output $\bar y(t)$ sampled at time $t=t_j$. Note that $\bar{y}(t)$ and $\bar u(t)$ are piecewise continuous.

{\blue
When we seek the values of $\alpha_f(t_j),\hat\theta(t_j),\hat{x}(t_j),\bar y(t_j))$ in the second equation of (\ref{eq:bar_y_and_u}), it typically requires the integration of the dynamics in (\ref{eq:xt}). For instance, $\hat x(t)=\hat x(t_0)+\int_{t_0}^t \chi(\hat x(\tau),\bar y(\tau))\mathrm d\tau $, $\hat\theta (t)=\hat\theta(t_0)+\int_{t_0}^t \vartheta(\hat\theta(\tau),\hat x(\tau),\bar y(\tau))\mathrm d\tau$  and $\alpha_f(t)=\alpha_f(t_0)+\int_{t_0}^t \varsigma (\alpha_f(\tau),\hat\theta(\tau),\hat x(\tau),\bar y(\tau))\mathrm d\tau$ where we replace $y(t)$ as $\bar y(t)$ due to \eqref{eq:bar_y_and_u}.
In order to further reduce the computational burden, we use sampled value of $\hat{x},\hat\theta,\alpha_f$ to construct the controller dynamics.
The controller in the paper is given in the form of
\begin{equation}
\begin{aligned}
\dot {\hat{x}}(t)&=\chi(\hat{x}(t_j),\bar y(t_j))\\
\dot {\hat{\theta}}(t)&=\vartheta(\hat\theta(t_j),\hat{x}(t_j),\bar y(t_j)) \\
\dot { \alpha}_f(t)&=\varsigma( \alpha_f(t_j),\hat\theta(t_j),\hat{x}(t_j),\bar y(t_j))\\
 u(t)&=\kappa( \alpha_f(t_j),\hat\theta(t_j),\hat{x}(t_j),\bar y(t_j))
\end{aligned}\label{eq:sample}
\end{equation}
for $t\in [t_j,t_{j+1})$ and $j\in \mathbb N$.
Note that  although a more general sampling scheduler can be used,   we choose to trigger sampling simultaneously with ED2 for the sake of  presentation simplicity.
The benefit of  (\ref{eq:sample}) over (\ref{eq:xt}) and (\ref{eq:u}) is that the value $\hat{x},\hat\theta,\alpha_f$ can be calculated by simple algebraic equation rather than numerical integration of the vector function. For instance,
 \begin{equation}
\hat{x}(t)=\hat{x}(t_j)+\chi(\hat{x}(t_j),\bar y(t_j)) (t-t_j) ,\quad  t\in[t_j,t_{j+1})
\end{equation}
where $\hat{x}(t_j)$ and  $\chi(\hat{x}(t_j),\bar y(t_j))$ are constant. The controller (\ref{eq:sample}) can also be implemented by an analog integration circuit rather  than on a computing unit. As will be explained later, this mechanism 
can be exploited to
 directly estimate next  triggering instance of some event conditions and thus simplify the event check.}

\brem
In \cite{zhang2021adaptive03}, the information transmitted between the plant and controller sides is also event-triggered. {\blue Although the state observer and adaptation dynamics  reside at the controller side, the
  dynamic filter   resides at the plant side to consume the continuous output signal $y(t)$, where the computation capacity is demanded.}
In this paper,
we move all components of the  controller to the controller side, but it brings a challenge that the virtual input designed in the backstepping technique
 is no longer first-order differentiable.   We will tackle this issue in the next section.
\erem

The objective is to design the controller and  event-triggered laws at the plant and controller sides such that
 all signals of the closed-loop system are semiglobally bounded and the output $y$ is practically stabilized to the origin, i.e., there exists a $\epsilon>0$ such that $\lim_{t\rightarrow \infty} |y(t)|\leq \epsilon$.



\section{Main Results \label{sec:MR}}
In this paper, the triggering law of  ED1 is designed as
\begin{equation}
\bar t_{k+1}=\inf\left\{t>\bar t_k:e_y\geq \gamma_y\right \}\label{eq4t}
\end{equation}
where  $e_y=\vert y(t)-y(\bar t_{k})\vert$ is the sampling error of output and $\gamma_y>0$ is a design parameter.

As will be elaborated in Section \ref{sub:ED}, the state of the estimator is decomposed as $\hat{x}=\xi+\theta \zeta$. Define $e_{\bar y} (t)=\vert \bar y(t)-\bar y(t_j)\vert$, $e_\xi(t)=\Vert \xi(t)-\xi(t_j)\Vert$, $e_\zeta(t)=\Vert \zeta(t)-\zeta(t_j)\Vert$, $e_h(t)=\vert\hat\theta(t)-\hat\theta(t_j)\vert$ and $e_f(t)=\Vert{\alpha}_{f}(t)-{\alpha}_{f}(t_j)\Vert$.
Let $\gamma_{\bar y}> \gamma_y$,
$\gamma_\xi,\gamma_\zeta,\gamma_h,\gamma_f>0$ be design parameters for ED2
 where $\gamma_y$ is given in (\ref{eq4t}) for  ED1.
 Then, the triggering law ED2 is designed as
\begin{align}
t_{j+1}=\inf \{&t>t_j:(e_{\bar y}\geq \gamma_{\bar y})\vee (e_\xi\geq \gamma_\xi)\vee (e_\zeta\geq \gamma_\zeta)
\notag \\
& \vee (e_h\geq \gamma_h) \vee (e_f\geq \gamma_f)
 \}\label{eq94}
\end{align}
{\blue where $\vee$ stands for `logic or'.}
In what follows, we will elaborate the design of the state observer, adaptive  law and stabilization controller, and specify  parameters in event-triggered laws.

\subsection{State Observer}\label{sub:ED}
Motivated by \cite{xing2018event02,krstic1995nonlinear06}, we  estimate the full state through two dynamics. One is  for the part of the plant that does not contain $\theta$ and its state is denoted as $\xi$, and the other one is for its unknown part and its state is denoted as $\zeta$.   $\xi$- and $\zeta$-dynamics are designed  as
\begin{align}
\dot{\xi}(t)&=A_c \xi(t_j)+k\bar{y}(t_j)+bu\label{eq92}\\
\dot{\zeta}(t)&=A_c\zeta(t_j)+\psi(\bar{y}(t_j)), \;\forall t\in[t_j,t_{j+1}),\ j\in\mathbb{N}\label{eq93}
\end{align}
where $\xi =[\xi_1,\cdots,\xi_n]\t$, $\zeta=[\zeta_c,\cdots,\zeta_n]\t$,
$b=[0,\cdots,1]\t$, {\blue $\psi(\bar y(t_j))=[\psi_1(\bar y(t_j)),\cdots,\psi_n(\bar y(t_j))]\t$} and $k=[k_1,\cdots,k_n]\t$ is chosen so that the matrix
\begin{equation}\label{eq10}
A_c=\left[\begin{array}{cccc}
-k_1&1&\cdots& 0\\
\vdots&\vdots&\ddots&\vdots\\
-k_{n-1}&0&\cdots&1\\
-k_n&0&\cdots&0
\end{array}\right]\in \mathbb R^{n\times n}
\end{equation}
is Hurwitz.

{\blue Now we explain the benefit of using the sampled value in the $\xi$-, $\zeta$-, $\hat\theta$- and $\alpha_f$-dynamics by examining the last four conditions in (\ref{eq94}).
Let $
 t_{\zeta,j+1}= \inf \{t>t_j:   e_\zeta\geq \gamma_\zeta
\}
$, $
 t_{\xi,j+1}= \inf \{t>t_j:   e_\xi\geq \gamma_\xi
\}
$,
$
 t_{h,j+1}= \inf \{t>t_j:   e_h\geq \gamma_h
\}
$
and
$
 t_{f,j+1}= \inf \{t>t_j:   e_f\geq \gamma_f
\}
$.
We take  $
	 t_{\zeta,j+1}
$ as the example. Note that $e_{\zeta}\geq\gamma_{\zeta}$
is equivalent to
$e_{\zeta}=\|\zeta(t)-\zeta(t_{j})\|=\|(A_{c}\zeta(t_{j})+\psi(\bar{y}(t_{j}))(t-t_{j})\|=\| (A_{c}\zeta(t_{j})+\psi(\bar{y}(t_{j}))\|(t-t_j)\geq\gamma_{\zeta}$, then $t_{\zeta,j+1}$ can be calculated at $t_j$
\begin{equation}
t_{\zeta,j+1}=t_j+\gamma_\zeta /\|(A_{c}\zeta(t_{j})+\psi(\bar{y}(t_{j}))\|\notag
\end{equation}
where we note that $\|(A_{c}\zeta(t_{j})+\psi(\bar{y}(t_{j}))\|$ is a known constant.
This idea also applies to calculating $
t_{\xi,j+1}$, $
t_{h,j+1}$ and $t_{f,j+1}$.
Let $\tilde t_{j+1}=\min \{t_{\zeta,j+1}, t_{\xi,j+1},t_{h,j+1},t_{f,j+1}\}$. Then, triggering condition in (\ref{eq94}) can be simplified as
$ t_{j+1}= \inf \{t>t_j:   (e_{\bar y}\geq \gamma_{\bar y}) \vee  (t\geq \tilde t_{j+1})
\}
$. Note that the condition $e_{\bar y}\geq \gamma_{\bar y}$ only needs to be checked when  the new transmitted output arrives. That means we only check condition  $e_{\bar y}\geq \gamma_{\bar y}$  at some discrete time instances when  $t_j<t< \tilde t_{j+1}$. If it is not satisfied for $t_j<t< \tilde t_{j+1}$, $t_{j+1}=\tilde t_{j+1}$. Therefore, continuously checking the event condition (\ref{eq94}) is avoided.}


The next lemma gives the error bound between actual output $y(t)$ and  sampled transmitted output $\bar y(t_j)$, {\blue whose   proof is given in appendix. }
\blem\label{lem1}
Under event-triggered laws (\ref{eq4t}) and (\ref{eq94}),  the inequality
\begin{equation}\label{eq158}
|y(t)-\bar y(t_j)|\leq \tilde\gamma_y,\forall t\in[t_j,t_{j+1}),\ j\in\mathbb{N}
\end{equation}
holds with $\tilde\gamma_y=\gamma_y+\gamma_{\bar y}$.
\elem

\emph {Unless otherwise specified, the rest of the analysis is based on time interval $t\in[t_j,t_{j+1})$. }We define
the  estimation error $\varepsilon$  as
\begin{equation}\label{eq5}
	\varepsilon=x-(\xi+\theta\zeta).
\end{equation}
{\blue The $x$-dynamics in \eqref{eq1} can be written in a compact form as
\begin{equation}\label{x-dynamics}
\dot x = Ax+\theta \psi (y)+bu
\end{equation}
where $\psi (y)=[\psi_1 (y),\cdots,\psi_n(y)]\t$, $b$ is denoted in \eqref{eq92} and
\begin{equation}
A=\left[\begin{array}{cc}0_{n-1}&I_{n-1}\\0&0_{n-1}\t\end{array}\right]\in \mathbb{R}^{n\times n}.
\end{equation}
By (\ref{eq92}), (\ref{eq93}), (\ref{eq5}) and \eqref{x-dynamics}, one has
\begin{align} \label{eq14}
\dot\varepsilon&= Ax+\theta \psi (y)-A_c \xi(t_j)-k\bar y(t_j)-\theta A_c \zeta(t_j)-\theta \psi(\bar y(t_j))\notag\\
& = Ax-ky -A_c \xi-\theta A_c \zeta +\theta\left [\psi(y)-\psi(\bar y(t_j))\right ]\notag\\
&\quad +k(y-\bar{y}(t_j))+A_c(\xi-\xi(t_j))+\theta A_c(\zeta-\zeta(t_j))\notag\\
&=A_c\varepsilon+k(y-\bar{y}(t_j))+\theta\left [\psi(y)-\psi(\bar y(t_j))\right ]\notag\\
&\quad +A_c(\xi-\xi(t_j))+\theta A_c(\zeta-\zeta(t_j))
\end{align}
where we used the fact $Ax-ky=A_c x$.} Define
\begin{equation}\label{eq16}
V_{\varepsilon}(\varepsilon)=\varepsilon\t P\varepsilon
\end{equation}
where $P$ is a positive definite matrix satisfying $PA_c+A_c\t P=-I$. The derivative of $V_\varepsilon$ along dynamics (\ref{eq14}) is
\begin{align}\label{eq17}
\dot{V}_{\varepsilon} = &-\Vert \varepsilon\Vert^2
+2\varepsilon\t Pk(y-\bar{y}(t_j))
+2\varepsilon\t PA_c(\xi-\xi(t_j))\notag\\
&+2\varepsilon\t P\theta\left [\psi(y)-\psi(\bar y(t_j))\right]+2\varepsilon\t P\theta A_c(\zeta-\zeta(t_j)).\notag
\end{align}
Using (\ref{eq94}), (\ref{eq158}) and  Young's inequality leads to
\begin{equation}\label{eq23}
	\dot{V}_{\varepsilon}\leq -(1-4\sigma)\Vert\varepsilon\Vert^2+d_\varepsilon
\end{equation}
with
\begin{align}\label{eq95}
d_\varepsilon=\Vert P\Vert^2\left[(\Vert k\Vert ^2+\bar\theta^2L^2)\tilde\gamma_y^2+(\gamma_\xi^2+\bar\theta^2\gamma_\zeta^2)\Vert A_c\Vert^2\right]/\sigma\notag
\end{align}
where $\sigma>0$ is a design parameter and $L=\sqrt{L_1^2+\cdots+L_n^2}$.

Next, we analyze the property of $\zeta$.
Define
\begin{equation}
V_{\zeta}(\zeta)=\zeta\t P\zeta.\notag
\end{equation}
Then the time derivative of $V_\zeta$ along dynamics (\ref{eq93}) is
\begin{equation}\label{eq40}
\dot V_{\zeta}=-\Vert \zeta\Vert^2+2\zeta\t PA_c( \zeta(t_j)-\zeta )+2\zeta\t P\psi(\bar{y}(t_j)).\notag
\end{equation}
Note that
$
	2\zeta\t P\psi(\bar{y}(t_j))
	\leq2\zeta\t P(\psi(\bar y(t_j))-\psi(0)+\psi(0))\leq2\Vert \zeta\Vert\Vert P\Vert (L\vert{y}\vert+L\tilde\gamma_y+\Vert \psi(0)\Vert)\leq2\sigma \Vert \zeta\Vert^2+a_\zeta y^2+d_{\zeta,1}
$
%
%
where $a_\zeta=\frac{1}{\sigma}\Vert P\Vert^2L^2$ is a known parameter and $d_{\zeta,1}=\frac{1}{\sigma}\Vert P\Vert^2(L\tilde\gamma_y+\Vert\psi(0)\Vert)^2$. Then, the bound of $\dot{V}_\zeta$ becomes
\begin{equation}\label{eq42}
\dot{V}_\zeta\leq -(1-3\sigma)\Vert \zeta\Vert^2+a_{\zeta}y^2+d_\zeta
\end{equation}
where $d_\zeta=d_{\zeta,1}+\frac{1}{\sigma}\Vert P\Vert^2\Vert A_c\Vert^2\gamma_\zeta^2$.

{\blue \brem\label{rem2}
The work \cite{zhang2021adaptive03}  uses $\dot{\hat x}_i=\hat x_{i+1}-k_i(\hat y-\bar y)$ for $i=1,\cdots,n-1$ and $\dot{\hat x}_n=u-k_n(\hat y-\bar y)$ as the
 state observer for system (\ref{eq1}).
 The estimation error is defined as $\bar \varepsilon$-dynamics with $\bar \varepsilon=x-\mbox{col}(\hat x_1,\cdots,\hat x_n)$ and its dynamics is rendered  to admit an input-to-state stable Lyapunov function,
 $\underline \alpha(\|\bar \varepsilon\|)\leq V(\bar \varepsilon)\leq \bar \alpha(\|\bar \varepsilon\|)$, $\dot V(\bar \varepsilon)\leq -\gamma V(\bar \varepsilon) +\Theta y^2+\tilde \Theta \|\varphi(0)\|^2+d_\Delta$, where $\gamma>0$,
$\varphi(0)=[\varphi_1(0),\cdots,\varphi_n(t_0)]\t$,
$\Theta$, $\tilde \Theta $ are unknown parameter relate to $\theta $, and $d_\Delta$ is a finite positive constant related to triggering threshold.
 In order to eliminate  the term $\Theta y^2$, the virtual controller to be designed needs to introduce another adaptive law to  estimate   $\Theta$, i.e, in the first step of backstepping procedures. And the virtual controller $\alpha_1$ is designed as $\alpha_1 =-c_1 \bar y -\hat\theta \psi_1 (\bar y)-\hat\Theta \bar y$ where $\hat \theta $ and $\hat\Theta$  denote the estimation of $\theta $ and $\Theta$, respectively.
 On the other hand, we avoid to use the high gain proposed in the state observer in \cite{qian2002output25,ahrens2009high26}.
\erem }

\subsection{Adaptive Law and Dynamic Filter}
The adaptive law is designed as
\begin{equation}
\dot{\hat{\theta}}(t)=\bar y(t_j)\left [\psi_1(\bar y(t_j))+\zeta_2(t_j) \right ]-\delta\hat\theta(t_j)\label{eq91}
\end{equation}
 for $\forall t\in[t_j,t_{j+1}),\ j\in\mathbb{N}$ where
$\hat \theta$ is the estimate of unknown parameter $\theta$ and
 $\delta>0$ is a design parameter.
 Let $\xi_i$, $\zeta_i$ and $\varepsilon_i$ be $i$th  element of $\xi$, $\zeta$ and $\varepsilon$ in (\ref{eq5}), respectively.
 We adopt dynamic filtering technique in \cite{swaroop2000dynamic09} and design the dynamic filter as
 \begin{equation}
 	\dot{\alpha}_{if}=\rho_i(-\alpha_{if}(t_j)+\alpha_{i-1}(t_j)),\; i=2,\cdots,n\label{eq29}
 \end{equation}
 where $\rho_i>0$ is design parameter. $\alpha_{i-1}$ is the virtual input   designed as
 \begin{align}
 	\alpha_1&=-c_1\bar{y}-\hat{\theta}\bar{\varpi}\label{eq32}\\
 	\alpha_i&=-c_i z_i-k_i(\bar y-\xi_1)-\rho_i \upsilon_i,\;i=2,\cdots,n-1 \label{eq81}
 \end{align}
 where $\bar\varpi=\psi_1(\bar y)+\zeta_2$ and $\hat \theta$ is given in adaptive law (\ref{eq91}).
  $\upsilon_i$ in (\ref{eq81}) is  an intermediate variable defined as
 \begin{align}
 \upsilon_i&=\alpha_{if}-\alpha_{i-1}, \;i=2,\cdots,n\label{eq30}\\
z_1&=y,\quad  z_i=\xi_i-\alpha_{if},\ i=2,\cdots,n\label{eq24}
\end{align}
where $z_i$ is the virtual error. 
We now recursively analyze the property of the virtual error.

\textbf{\emph{Step 1}}:
By (\ref{eq5}), (\ref{eq24}) and (\ref{eq30}), we have $x_2=\xi_2+\theta\zeta_2+\varepsilon_2$, $\xi_2=z_2+\upsilon_2+\alpha_1$. Then, the derivative of $z_1$ is
\begin{equation}\label{eq25}
\dot{z}_1=x_2+\theta\psi_1(y)=z_2+\upsilon_2+\alpha_1+\theta\varpi+\varepsilon_2
\end{equation}
with $\varpi=\psi_1(y)+\zeta_2$.
  Backstepping method with dynamic filter requires the virtual input   $\alpha_i$ be at least  first-order differentiable, however  $\alpha_1$ in (\ref{eq32}) is piecewise continuous.  In order to tackle this issue, we define companion variables $\hat\alpha_1$, $\hat \upsilon_2$ for $\alpha_1$, $ \upsilon_2$, respectively, as 
\begin{equation}\label{eq78}
\hat\alpha_1=-c_1y-\hat\theta \varpi,\; \hat\upsilon_2=\alpha_{2f}-\hat\alpha_1.
\end{equation}
By noting   $\upsilon_2+\alpha_1=\hat\upsilon_2+\hat\alpha_1$,  (\ref{eq25}) and (\ref{eq78}), one has
\begin{align}
	z_1\dot{z}_1  &=  z_1(z_2+\hat\upsilon_2+\hat\alpha_1+\theta\varpi+\varepsilon_2)\notag\\
	  &= z_1(-c_1z_1+z_2+\upsilon_2+\varepsilon_2)+\tilde\theta z_1\varpi\notag
\end{align}
where
 \begin{equation}
 \tilde\theta=\theta-\hat\theta\notag
 \end{equation}
 denotes the parameter estimate error. Define
 \begin{equation}
 V_1(z_1,\hat\upsilon_2,\tilde\theta,\varepsilon,\zeta) =\frac{1}{2}(z_1^2+\tilde{\theta}^2+\hat\upsilon_2^2)+V_{\varepsilon}+V_{\zeta}.\notag
 \end{equation}
Then $\dot V_1$ is
\begin{align}\label{eq57}
\dot V_1&=z_1(-c_1z_1+z_2+\hat\upsilon_2+\varepsilon_2)+\hat\upsilon_2\dot{\hat\upsilon}_2+\dot V_\varepsilon+\dot V_\zeta\notag\\
&\quad +\tilde\theta(\tau-\dot{\hat\theta})+\tilde\theta(z_1\varpi-\tau).
\end{align}
where $\tau=\bar y\big(\psi_1(\bar y)+\zeta_2\big)-\delta \hat\theta$.
Now, we first examine
\begin{align}
	\tilde{\theta}(\tau-\dot{\hat{\theta}})&=\tilde{\theta} [\bar y \psi_1(\bar y)-\bar y(t_j)\psi_1(\bar y(t_j))  \nonumber\\
&\quad +  \bar y\zeta_2 -\bar y(t_j) \zeta_2(t_j) -\delta \hat\theta +\delta\hat\theta(t_j)].\label{eq:tau_error}
\end{align}
Note that
\begin{align}\label{eq46}
&\tilde{\theta}\left [\bar{y}\psi_1(\bar{y})-\bar{y}(t_j)\psi_1(\bar{y}(t_j))\right]\notag\\
&=\tilde{\theta}\bar y \left [ \psi_1(\bar y)-\psi_1(\bar y(t_j))\right]+\tilde{\theta}(\bar y-\bar{y}(t_j)) \psi_1(\bar y(t_j)).
\end{align}
From (\ref{eq2}) and (\ref{eq4t}), one has
\begin{align}\label{eq47}
&\tilde \theta\bar y\left[ \psi_1(\bar y)-\psi_1(\bar y(t_j)) \right ]\notag\\
&\leq \vert \tilde\theta\vert (\vert y\vert+\tilde\gamma_y)L\tilde\gamma_y\leq2\delta_1\tilde\theta^2 +a_{1,1}y^2+d_{1,1}
\end{align}
and
\begin{align}\label{eq48}
\tilde\theta( \bar y-\bar y(t_j) )\psi_1(\bar y(t_j))&\leq \vert \tilde\theta\vert\gamma_{\bar y} (L\vert y\vert+L\tilde\gamma_y+\vert \psi_1(0)\vert)\notag\\
&\leq  2\delta_1\tilde\theta^2
+a_{1,1}y^2+d_{1,1}
\end{align}
where $\delta_1=1/(20\sigma)$, $a_{1,1}=5\sigma L^2\tilde\gamma_y^2$ and $d_{1,1}=5\sigma\tilde\gamma_y^2(L\tilde\gamma_y+\vert\psi_1(0)\vert)^2$ with a design parameter   $\sigma>0$. From (\ref{eq46}),
(\ref{eq47})
and  (\ref{eq48}), one has
\begin{equation}\label{eq49}
\tilde{\theta} \left[\bar{y}\psi_1(\bar{y})-\bar{y}(t_j)\psi_1(\bar{y}(t_j))\right]\leq 4\delta_1\tilde\theta^2 +2a_{1,1}y^2+2d_{1,1}.
\end{equation}
Similar to (\ref{eq47})-(\ref{eq49}), one has
\begin{equation}\label{eq50}
\tilde\theta\left[\bar y\zeta_2-\bar y(t_j)\zeta_2(t_j)\right]\leq \delta_2\tilde\theta^2 +\sigma \zeta_2^2+a_{1,2}y^2+2d_{1,2}
\end{equation}
where $\delta_2=(3 +5\tilde \gamma_y^2)/(20\sigma)$, $a_{1,2}=5\sigma\gamma_{\zeta}^2$ and $d_{1,2}=5\sigma\tilde\gamma_y^2\gamma_{\zeta}^2$.  Moreover,  one has
\begin{equation}\label{eq52}
\delta\tilde\theta\big(\hat\theta(t_j)-\hat{\theta}\big)\leq \vert \tilde\theta\vert\delta\gamma_h\leq \delta_1\tilde\theta^2+d_{1,3}
\end{equation}
where $d_{1,3}=5\sigma\delta^2\gamma_h^2$.
Using (\ref{eq:tau_error}), (\ref{eq49}), (\ref{eq50}) and (\ref{eq52}) obtains
\begin{align}\label{eq135}
\tilde\theta(\tau-\dot{\hat{\theta}})&\leq(5\delta_1+\delta_2)\tilde\theta^2 +(2a_{1,1}+a_{1,2})y^2+\sigma\zeta_2^2\notag\\
&\quad +2d_{1,1}+2d_{1,2}+d_{1,3}.
\end{align}
Next, we examine $\tilde\theta\big(z_1 \varpi-\tau\big)$. Due to
\begin{align*}
	&\tilde\theta(y\psi_1(y)-\bar y\psi_1(\bar y))
	\leq 2\delta_1\tilde\theta^2 +4a_{1,1}y^2 +2d_{1,1}\\
	&\tilde\theta(y\zeta_2-\bar y\zeta_2)\leq \delta_3\tilde\theta^2+\sigma\zeta_2^2\\
	&\delta\tilde\theta\hat\theta=\delta\tilde\theta(\theta-\tilde\theta)\leq -\delta\tilde\theta^2/2+d_{1,4}
\end{align*}
where $\delta_3=\tilde\gamma_y^2/(4\sigma)$ and $d_{1,4}=\delta\bar\theta^2/2$, one has
\begin{align}\label{eq80}
\tilde\theta(z_1\varpi-\tau)&\leq (-\delta/2+2\delta_1+\delta_3)\tilde\theta^2 +4a_{1,1} y^2+\sigma \zeta_2^2\notag\\
& \quad +2d_{1,1}+d_{1,4}.
\end{align}
Due to
\begin{gather}\label{eq99}
\dot{\hat\alpha}_1 = \frac{\partial \hat \alpha_1}{\partial y}(x_2+\theta\psi_2)+\frac{\partial \hat  \alpha_1}{\partial \hat\theta}\left[\bar y(t_j)(\psi_1(\bar y(t_j))+\zeta_2(t_j))\right.\notag\\
 -\delta\hat\theta(t_j)\big ]+\frac{\partial \hat  \alpha_1}{\partial \zeta_2}\left[\zeta_3(t_j)-k_2\zeta_1(t_j)+\psi_2(\bar y(t_j))\right]\notag
\end{gather}
which means that $ {\hat \alpha}_1$ is first-order differentiable and $\dot {\hat \alpha}_1$ is piecewise continuous.
Let $\varrho_2$ and $\phi_2$ are positive  parameters to be designed later and $\rho_2$ is selected such that  $\rho_2\geq 2+\phi_2+\varrho_2$. Then
\begin{align}\label{eq43}
\hat\upsilon_2\dot{\hat\upsilon}_2&=\hat\upsilon_2(\rho_2(-\alpha_{2f}(t_j)+\alpha_1(t_j))-\dot{\hat\alpha}_1    )\notag\\
&=-\rho_2\hat\upsilon_2^2+\hat\upsilon_2\rho_2(\alpha_{2f}-\alpha_{2f}(t_j)) \notag\\
&\quad +\hat\upsilon_2\rho_2(\alpha_1(t_j)-\alpha_1)+\hat\upsilon_2\rho_2(\alpha_1-\hat\alpha_1)-\hat\upsilon_2\dot{\hat \alpha}_1\notag\\
&\leq -(\varrho_2+3/2)\hat\upsilon_2^2+\hat\upsilon_2\rho_2(\alpha_1(t_j)-\alpha_1)\notag\\
&\quad +\hat\upsilon_2\rho_2(\alpha_1-\hat\alpha_1)+\dot{\hat\alpha}_1^2/(2\phi_2)+\rho_2^2\gamma_f^2/2.
\end{align}
Combining (\ref{eq23}), (\ref{eq42}), (\ref{eq57}), (\ref{eq135}), (\ref{eq80}) and (\ref{eq43}), one has
\begin{align}\label{eq58}
\dot V_1&\leq z_2^2/2-\bar c_1z_1^2-(\varrho_2+1)\hat\upsilon_2^2-\bar\delta\tilde\theta^2-(1-5\sigma)\Vert\varepsilon\Vert^2\notag\\
&\quad -(1-5\sigma)\Vert \zeta\Vert^2+\hat\upsilon_2\rho_2(\alpha_1(t_j)-\alpha_1)\notag\\
&\quad +\hat\upsilon_2\rho_2(\alpha_1-\hat\alpha_1)+\dot{\hat\alpha}_1^2/(2\phi_2)+d_1
\end{align}
with
\begin{equation}\label{eq59}
\begin{aligned}
\bar c_1&=c_1-1-1/(4\sigma)-6a_{1,1}-a_{1,2}-a_{\zeta}\\
\bar\delta&=\delta/2-7\delta_1-\delta_2-\delta_3\\
d_1&=4d_{1,1}+2d_{1,2}+d_{1,3}+d_{1,4}+\rho_2^2\gamma_f^2/2+d_\varepsilon+d_\zeta.
\end{aligned}
\end{equation}


\textbf{\emph{Step $\boldsymbol{i,\ i=2,\cdots,n-1}$}}:
 let $\hat\alpha_i$, $\hat\upsilon_i$ be  companion variables for  $\alpha_i$, $\upsilon_i$, respectively and defined as
 \begin{equation}\label{eq62}
\begin{gathered}
	\hat\alpha_i=-c_iz_i-k_i\big( y-\xi_1\big)-\rho_i\hat\upsilon_i\\
	\hat\upsilon_i=\alpha_{if}-\hat\alpha_{i-1},\quad  i=2,\cdots,n-1.
\end{gathered}
\end{equation}
Then, the derivative of $z_i$ is
\begin{align}\label{eq60z}
\dot z_i&=\xi_{i+1}(t_j)+k_i(\bar y(t_j)-\xi_1(t_j))+\rho_i(\alpha_{if}(t_j)-\alpha_{i-1}(t_j))\notag\\
&=z_{i+1}+\hat\upsilon_{i+1}+\hat\alpha_{i}+(\xi_{i+1}(t_j)-\xi_{i+1})+\rho_i\hat\upsilon_i\notag\\
&\quad +k_i(\bar y(t_j)-\xi_1(t_j))+\rho_i(\alpha_{if}(t_j)-\alpha_{if})\notag\\
&\quad +\rho_i(\hat\alpha_{i-1}-\alpha_{i-1})+\rho_i(\alpha_{i-1}-\alpha_{i-1}(t_j)).\notag
\end{align}
Define
\begin{equation}\label{eq61}
V_i (z_i,\hat\upsilon_{i+1})=\frac{1}{2}(z_i^2+\hat\upsilon_{i+1}^2).\notag
\end{equation}
Similar to step 1, substituting (\ref{eq62}) into the derivative of $V_i$  gives
\begin{gather}
\dot V_i\leq z_{i+1}^2/2- (\bar c_i+3/2)z_i^2+\hat\upsilon_{(i+1)f}\rho_{i+1}(\alpha_i(t_j)-\alpha_i)\notag\\
-(\varrho_i+1)\upsilon_{i+1}^2+\hat\upsilon_{(i+1)f}\rho_{i+1}(\alpha_i-\hat\alpha_i)+\dot{\hat\alpha}_{i}^2/(2\phi_i)\notag\\
+z_i\rho_i(\hat\alpha_{i-1}-\alpha_{i-1})+z_i\rho_i(\alpha_{i-1}-\alpha_{i-1}(t_j))+d_i\label{eq63}
\end{gather}
where  $\varrho_i$ and $\phi_i$ are positive  parameters to be designed later, $\rho_i$ is selected such that
 $\rho_i\geq 2+\phi_i+\varrho_i$ and 
 \begin{gather}\label{EQ10}
 	\bar c_i=c_i-9/2,\quad d_i=k_i^2(\tilde\gamma_y^2+\gamma_\xi^2)/2+\gamma_\xi^2/2+\rho_{i+1}^2\gamma_f^2/2.\notag
 	\end{gather}
Due to
\begin{gather}
\dot{\hat\alpha}_i=\sum_{p=1,i}\frac{\partial \hat\alpha_p}{\partial \xi_p}\left [\xi_{p+1}(t_j)+k_p(\bar y(t_j)-\xi_1(t_j))\right]+\frac{\partial \hat\alpha_i}{\partial \hat\alpha_{i-1}}\dot{\hat \alpha}_{i-1}\notag\\
+\frac{\partial\hat \alpha_i}{\partial \alpha_{if}}\rho_i(-\alpha_{if}(t_j)+\alpha_{i-1}(t_j))+\frac{\partial \hat\alpha_i}{\partial y}(x_2+\theta \psi_2)\label{eq100}
\end{gather}
and  recursive steps,
$\dot {\hat \alpha}_i$ is piecewise continuous and $ {\hat \alpha}_i$ is first-order differentiable.

\textbf{\emph{Step $\boldsymbol {n}$}}: Define $V_n (z_n)=z_n^2/2$,
then the derivative of $V_n$ is
\begin{equation}\label{eq64}
\dot V_n=z_n\left[u+k_n(\bar y(t_j)-\xi_1(t_j))+\rho_n\upsilon_n(t_j)\right]
\end{equation}
where we note $\upsilon_n(t_j)=\alpha_{nf}(t_j)-\alpha_{n-1}(t_j)$.
\subsection{Controller Design and Stability Analysis}
Finally, we
design the stabilization controller as
\begin{equation}
	u(t)=  -c_nz_n(t_j)-k_n( \bar y(t_j)-\xi_1(t_j))-\rho_n \upsilon_n(t_j)\label{eq86}
\end{equation}
 for $\forall t\in[t_j,t_{j+1})$. Substituting (\ref{eq86}) into (\ref{eq64}) yields
\begin{equation}\label{eq101}
\dot V_n=- c_nz_n^2+c_nz_n(z_n-z_n(t_j)).\notag
\end{equation}
{\blue
Denote
\begin{equation}\label{eq:eta}
\eta=[z\t,\hat\upsilon\t,\tilde \theta ,\varepsilon\t,\zeta\t]\t\in\mathbb R^{4n}
\end{equation}
with $z=[z_1,\cdots,z_n]\t$,  $\hat \upsilon=[\hat\upsilon_2,\cdots,\hat \upsilon_n]\t$. And denote
\begin{equation}\label{eq:beta}
\beta=[c\t,   \rho\t, \phi\t, \delta, \sigma]\t\in \mathbb R^{3n}
\end{equation}
 where $c=[c_1,\cdots,c_n]\t$ are feedback gains in
(\ref{eq32}) and (\ref{eq81}),  $\rho=[\rho_2,\cdots,\rho_n]\t$ are dynamic filter gains in (\ref{eq29}), $\delta$ is the   parameter of  adaptive law in (\ref{eq91}), and $\phi=[\phi_2,\cdots,\phi_n]\t, \sigma$ are adjustable parameters introduced in the recursive steps.}

We suppose  state observer gains
$k=[k_1,\cdots,k_n]\t$  in (\ref{eq10}) are determined before the selection of $\beta$.
Let $\gamma=[\gamma_y,\gamma_{\bar y}, \gamma_\xi,\gamma_\zeta,\gamma_f,\gamma_h]$.

Define the  Lyapunov function candidate
\begin{equation}\label{eq75}
V(\eta)=V_1+\cdots+V_n.
\end{equation}
Then, we have the following results.
\bthm\label{the1}
Consider the uncertain system (\ref{eq1}) satisfying Assumption \ref{ass1} and the  controller composed of
stabilization controller (\ref{eq86}), adaptive law (\ref{eq91}), state observer (\ref{eq92}), (\ref{eq93}), filter dynamics (\ref{eq29}) under triggering laws   (\ref{eq4t}) and (\ref{eq94}). Then
there exists a $\underline q$ such that  for any given positive number $q> \underline q $
there always exists a set of  parameters $\beta$ such that 
the set
{\blue$	\Omega_\eta(q)=\{\eta \in \mathbb{R}^{4n}\mid V(\eta)\leq q\}\label{eq:Omega}
$
}
 is a positively invariant set and
 the solution of the closed-loop system is semiglobally bounded. Moreover, the output $y$ is practically stabilized to the origin,  and the Zeno behavior is excluded. The procedures of the controller design are organized in Algorithm 1.
\ethm
\proofnow
Note that
within $\Omega_\eta(q)$,  variables $\vert z_i\vert, i=1,\cdots,n$, $\vert \hat\upsilon_i\vert, i=2,\cdots,n$, $\vert \tilde\theta\vert$, $\Vert \varepsilon\Vert$, $\Vert \zeta\Vert$ have upper bounds $\bar z_i$, $\hat\Upsilon_i$, $\tilde\Theta$, $\bar \varepsilon$, $\bar \zeta$, respectively. They  are   chosen as
\begin{equation}\label{eq149}
\bar z_i={\hat\Upsilon}_i=\tilde\Theta=\sqrt{2q},\quad  \bar\varepsilon=\bar \zeta=\sqrt{q/\underline{\lambda}}\notag
\end{equation}
where $\underline{\lambda}$ denotes the minimum eigenvalue of $P$.
The proof will be divided into three parts. First, we show  the  bound of $\alpha_i-\hat\alpha_{i},\ i=1,\cdots,n-1$ within  compact set $\Omega_\eta(q)$.  Note that
\begin{align}\label{eq104}
\vert \alpha_1-\hat\alpha_1\vert&\leq  c_1\vert\bar y- y\vert+\vert \hat\theta\vert \vert \psi_1(\bar y)-\psi_1(y)\vert\notag\\
&\leq  c_1\tilde\gamma_y+\hat\Theta L\tilde\gamma_y\triangleq \iota_1\notag
\end{align}
where $\hat\Theta=\bar\theta+\tilde\Theta$ and  
\begin{equation}\label{eq108}
\vert \hat\alpha_1-\hat\alpha_1(t_j)\vert\leq  c_1\vert \bar y(t_j)-\bar y\vert+\vert \hat\theta(t_j)\bar \varpi(t_j)-\hat\theta\bar\varpi\vert.\notag
\end{equation}
Due to
\begin{align}
&\vert \hat\theta(t_j)\bar \varpi(t_j)-\hat\theta\bar\varpi\vert\notag\\
&=\vert \hat\theta(t_j)\bar\varpi(t_j)-\hat\theta(t_j)\bar\varpi+\hat\theta(t_j)\bar\varpi-\hat\theta\bar\varpi\vert\notag\\
&\leq \vert \hat\theta(t_j)\vert \vert \bar\varpi(t_j)-\bar\varpi\vert+\vert \hat\theta-\hat\theta(t_j)\vert\vert \bar\varpi\vert\notag\\
&\leq (\gamma_h+\hat\Theta)(L\tilde\gamma_y+\gamma_\zeta)+\gamma_h(L\tilde\gamma_y+L\vert y\vert +\vert \zeta_2\vert)\notag\\
&\leq (\gamma_h+\hat\Theta)(L\tilde\gamma_y+\gamma_\zeta)+\gamma_h(L\tilde\gamma_y+L\bar z_1 +\bar \zeta)\notag
\end{align}
one has
\begin{align}\label{eq110}
\vert \hat\alpha_1-\hat\alpha_1(t_j)\vert&\leq c_1\tilde\gamma_y+(\gamma_h+\hat\Theta)(L\tilde\gamma_y+\gamma_\zeta)\notag\\
&\quad +\gamma_h(L\tilde\gamma_y+L\bar z_1 +\bar \zeta)\triangleq \Lambda_1.\notag
\end{align}
Since
$\alpha_2,\hat\alpha_2$ are defined in (\ref{eq81}), (\ref{eq62}), respectively, then we have $	\vert \alpha_2-\hat\alpha_2\vert\leq \vert k_2\vert \vert y-\bar y\vert+\rho_2\vert \alpha_1-\hat\alpha_1\vert
\leq \vert k_2\vert \tilde\gamma_y+\rho_2\iota_1\triangleq \iota_2$ and $	\vert \hat\alpha_2-\hat\alpha_2(y_j)\vert\leq (c_2+\rho_2)\vert \alpha_{2f}-\alpha_{2f}(t_j)\vert+c_2\vert \xi_2-\xi_2(t_j)\vert
+\vert k_2\vert (\vert y-y(t_j)\vert+\vert\xi_1-\xi_1(t_j)\vert)+\rho_2\vert \hat\alpha_1-\hat\alpha_1(t_j)\vert\leq (c_2+\rho_2)\gamma_f+c_2\gamma_\xi+\vert k_2\vert(\tilde\gamma_y+\gamma_\xi)+\rho_2\Lambda_1\triangleq \Lambda_2$.
Similarly, step by step, we can conclude the following inequalities
\begin{equation}\label{eq107}
\vert \alpha_i-\hat\alpha_i\vert\leq \iota_i,\ \vert \hat\alpha_i-\hat\alpha_i(t_j)\vert\leq \Lambda_i,\; i=1,\cdots,n-1
\end{equation}
for some $\iota_i, \Lambda_i>0$ depending on $c,k,\rho,\gamma,q$. 
 It's worth noting that when  all elements of $\gamma$ are chosen sufficiently small,    $\iota_i,\Lambda_i,i=1,\cdots,n$ are made sufficiently small. Therefore, by (\ref{eq107})
\begin{equation}
	\vert \upsilon_n\vert=\vert \hat\upsilon_n+\hat\alpha_{n-1}-\alpha_{n-1}\vert\leq{\hat\Upsilon}_n+\iota_{n-1}.\notag
	\end{equation}

Next, we will show that $\Omega_\eta(q)$ is a positively invariant set and  all internal signals in the closed-loop system is  semiglobal boundedness. By (\ref{eq5}), we have
\begin{equation}\label{eq119}
\vert \xi_1\vert=\vert y-\theta\zeta_1-\varepsilon_1\vert\leq \bar z_1+\vert\theta\vert\bar\zeta+\bar\varepsilon\triangleq \bar\xi_1.\notag
\end{equation}
By (\ref{eq107}), $u$ satisfies
\begin{align}\label{eq120}
\vert u\vert&\leq  c_n\vert z_n(t_j)\vert+\vert k_n\vert (\vert \bar y(t_j)\vert+\vert \xi_1(t_j)\vert)+\rho_n\vert \upsilon_n(t_j)\vert\notag\\
&\leq c_n\bar z_n+\vert k_n\vert(\bar z_1+\bar \zeta_1)+\rho_n({\hat\Upsilon}_n+\iota_{n-1})\triangleq\bar u.\notag
\end{align}
Let $
	V_\xi=\xi\t P\xi
$,
the derivative of $V_\xi$ along
$\xi$-dynamics in (\ref{eq92})
 is bounded as $\dot V_\xi\leq -(1-4\sigma)\bar \lambda ^{-1}V_\xi+d_\xi$, where $d_\xi=\Vert P\Vert^2(\Vert A_c\Vert^2\gamma_\xi^2+2L^2q+L^2\tilde\gamma_y^2+\Vert b\Vert^2\bar u^2)/\sigma$ where we used the fact that $\vert y\vert\leq\sqrt{2q}$ when states stay in the compact set $\Omega_\eta(q)$. Therefore,
the upper bound of $\xi$ is
\begin{equation}\label{eq123}
\Vert \xi\Vert \leq \sqrt{\bar\lambda(\Vert\xi(0)\Vert+ d_\xi)/(\underline\lambda(1-4\sigma))}\triangleq \bar\xi.
\end{equation}
Now, we   analyze the property of $\dot{\hat\alpha}_i,i=1,\cdots,n-1$ within compact set $\Omega_\eta(q)$. By (\ref{eq78}), the first partial derivatives of $\hat\alpha_1$ are
\begin{equation}\label{eq137}
\frac{\partial\hat \alpha_1}{\partial y}=-c_1-\hat\theta\frac{\partial \psi_1(y)}{\partial y},\quad \frac{\partial \hat\alpha_1}{\partial \hat\theta}=-\varpi,\quad \frac{\partial \hat\alpha_1}{\partial \zeta_2}=-\hat\theta.\notag
\end{equation}

Due to $\vert\hat\theta\vert\leq \vert\tilde\theta\vert+\vert \theta\vert\leq 
\hat\Theta$, $\vert\xi_i\vert\leq \Vert \xi\Vert\leq \bar\xi$, and $\vert \zeta_i\vert \leq\Vert \zeta\Vert\leq \bar\zeta$,  by (\ref{eq136}), we have
\begin{align}
\vert \dot {\hat \alpha}_1(t)\vert&\leq  (c_1+\hat\Theta\Psi_1)(\bar\xi+\bar\theta \bar\zeta+\bar\varepsilon+\bar\theta L\bar z_1) \notag\\
&\quad +(L\bar z_1+\bar \zeta)\big(\bar z_1(L\bar z_1+\bar \zeta)+\delta \hat\Theta\big)\notag\\
&\quad +\hat\Theta\big((1+\vert k_2\vert)\bar \zeta+L\bar z_1\big)\triangleq \Delta_1.\notag
\end{align}
 By (\ref{eq62}) and (\ref{eq100}), $\vert\dot{\hat\alpha}_2\vert$ satisfies
\begin{align}\label{eq139}
\vert\dot {\hat \alpha}_2(t)\vert&\leq  k_2(2+k_1+k_2)\bar \xi+k_2(k_1+k_2)\bar z_1+k_2(\bar\xi+\bar\theta \bar\zeta\notag\\
&\quad +\bar\varepsilon+\bar\theta L\bar z_1)+(c_2+\rho_2){\hat\Upsilon}_2+\rho_2\Delta_1\triangleq \Delta_2.\notag
\end{align}
 Similarly, we can  derive that $\vert\dot{\hat\alpha}_i\vert , i=3,\cdots,n-1$ is also  upper bounded, 
 that is
\begin{equation}\label{eq103}
\vert \dot {\hat \alpha}_i(t)\vert \leq \Delta_i , \ t\in[0,\infty)
\end{equation}
with $\Delta_i$ depending on
$c$, $k$, $\rho$, $\delta$, $q $,
   and increasing as each scalar variable of  $c$, $k$, $\rho$, $\delta$, $q $ increases.

\begin{algorithm}
\caption{Event-triggered adaptive output feedback control}
\label{tab:parametervalues}
\begin{algorithmic}[1]
\State
Design initial conditions $\xi(0)$, $\zeta(0)$, $\hat\theta(0)$, select suitable parameter $\beta$ such that $c\geq d/q$ is satisfied.
\State
Set ED1, ED2 as (\ref{eq4t}) and  (\ref{eq94}).
\State
Set  state observer (\ref{eq92}), (\ref{eq93}) based on the sampled value.
\State
Design the available virtual controller (\ref{eq32}) for step 1 with  event-triggered based adaptive law (\ref{eq91}).
\State
Design virtual controller (\ref{eq81}) with event-triggered based $\alpha_{if}$-dynamics (\ref{eq29}), form the stabilization controller (\ref{eq86}).
\end{algorithmic}
\end{algorithm}

Due to $z_n-z_n(t_j)=\xi_n-\xi_n(t_j)+\alpha_{nf}(t_j)-\alpha_{nf}$,  the bound of $\dot V_n$ becomes
\begin{equation}\label{eq117}
\dot V_n\leq -(\bar c_n+1/2)z_n^2+d_n
\end{equation}
where
\begin{gather}\label{EQ11}
	\bar c_n=c_n-1,\ d_n=c_n^2(\gamma_\xi+\gamma_f)^2/2.\notag
	\end{gather}
Due to (\ref{eq58}) and \eqref{eq63}, one has $\hat\upsilon_{(i+1)f}\rho_{i+1}(\alpha_i(t_j)-\alpha_i)\leq \frac{1}{2}\hat\upsilon_{(i+1)f}^2+\frac{1}{2}\rho_{i+1}^2\Lambda_i^2$, $\hat\upsilon_{(i+1)f}\rho_{i+1}(\alpha_i-\hat\alpha_i)\leq\frac{1}{2}\hat\upsilon_{(i+1)f}^2+\frac{1}{2}\rho_{i+1}^2\iota_i^2$, $z_i\rho_i(\hat\alpha_{i-1}-\alpha_{i-1})\leq \frac{1}{2}z_i^2+\frac{1}{2}\rho_i^2\iota_{i-1}^2$, $z_i\rho_i(\alpha_{i-1}-\alpha_{i-1}(t_j))\leq\frac{1}{2}z_i^2+\frac{1}{2}\rho_i^2\Lambda_{i-1}^2$.
\begin{figure*}
\centering
\begin{minipage}{0.3\linewidth}
		\centering
		\includegraphics[width=1\linewidth]{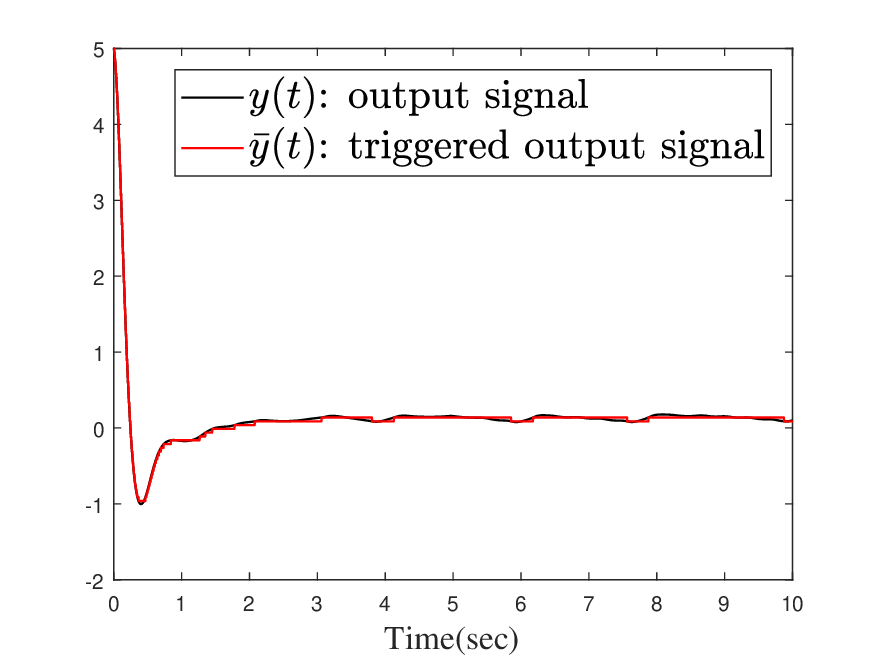}
\captionsetup{font={small}}
\vspace{-8pt}
		\caption{\small{The trajectories of actual output  $y(t)$ and triggered output $\bar y(t)$.}}
		\label{fig3}
	\end{minipage}
	\begin{minipage}{0.3\linewidth}
		\centering
		\includegraphics[width=1\linewidth]{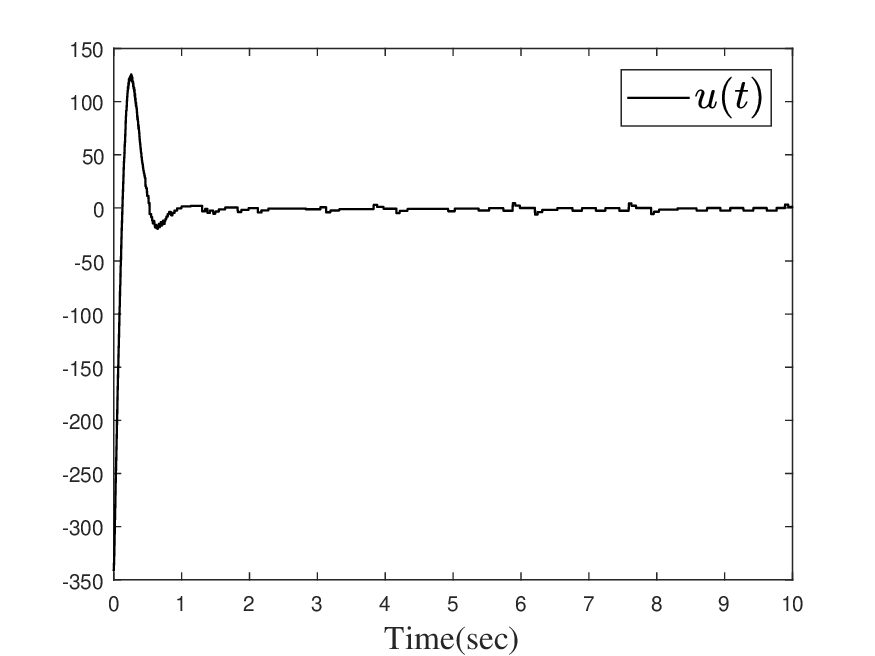}
\vspace{-8pt}
		\caption{\small{The trajectory of actual controller $u(t)$.}}
		\label{fig4}
	\end{minipage}
\begin{minipage}{0.3\linewidth}
		\centering
		\includegraphics[width=1\linewidth]{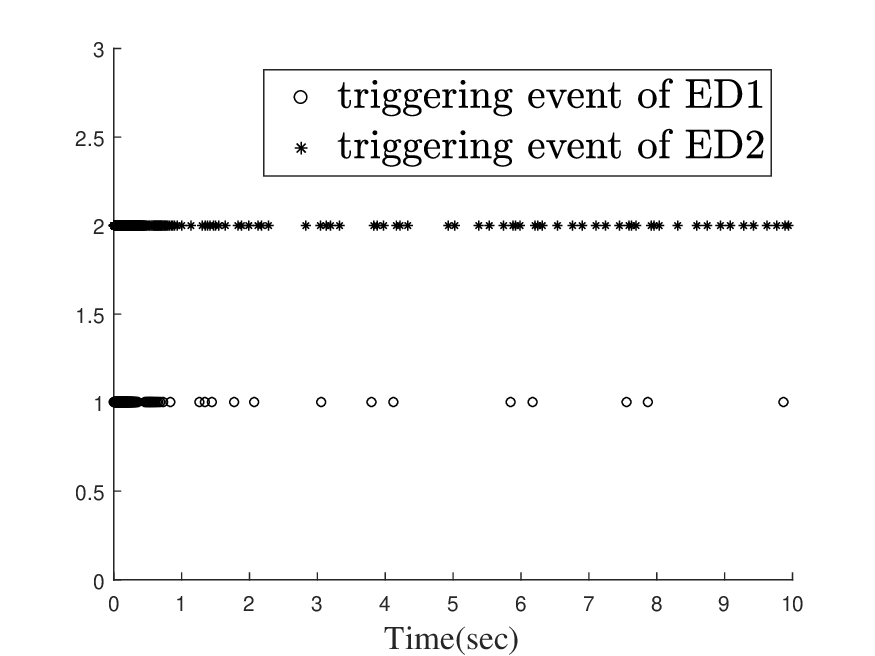}
\captionsetup{font={small}}
\vspace{-8pt}
		\caption{\small{Trigger events of ED1 and ED2.}}
		\label{fig8}
	\end{minipage}
\end{figure*}

\begin{figure*}
\centering
\begin{minipage}{0.3\linewidth}
		\centering
		\includegraphics[width=1\linewidth]{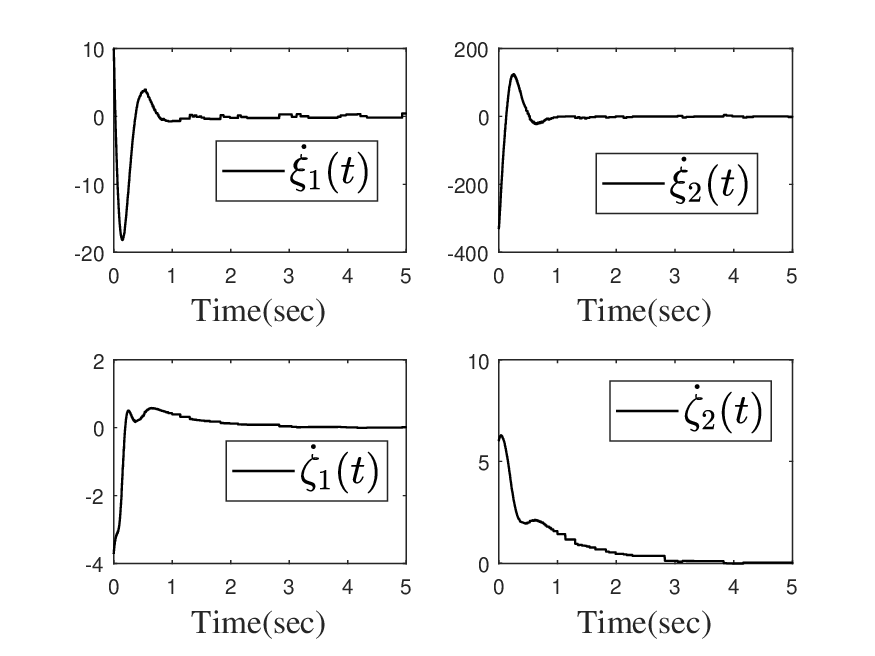}
\captionsetup{font={small}}
\vspace{-8pt}
		\caption{\small{The trajectories of state observer $\dot\xi_1(t)$, $\dot\xi_2(t)$, $\dot\zeta_1(t)$, $\dot\zeta_2(t)$.}}
		\label{fig5}
	\end{minipage}
	\begin{minipage}{0.3\linewidth}
		\centering
		\includegraphics[width=1\linewidth]{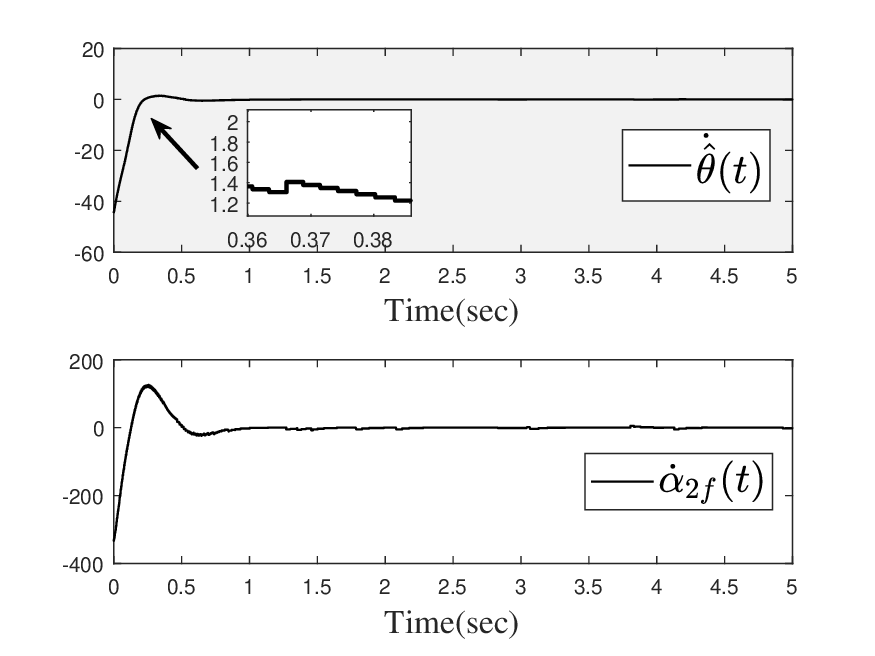}
\captionsetup{font={small}}
\vspace{-8pt}
		\caption{\small{The trajectories of adaptive law $\dot\theta(t)$ and $\alpha_{2f}$-dynamics.}}
		\label{fig6}
	\end{minipage}
\begin{minipage}{0.3\linewidth}
		\centering
		\includegraphics[width=1\linewidth]{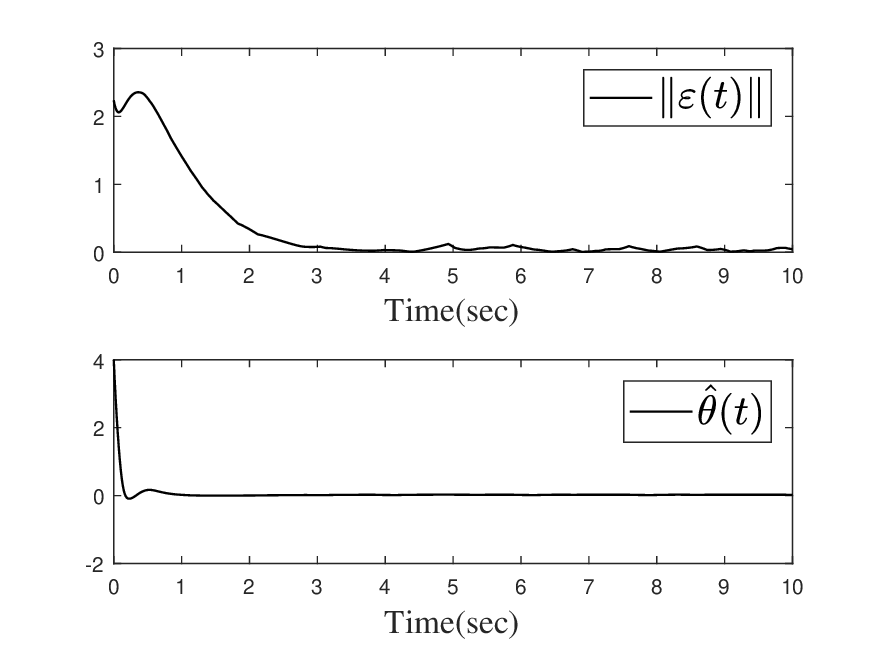}
\captionsetup{font={small}}
\vspace{-8pt}
		\caption{\small{The trajectories of $\varepsilon(t)$ and $\hat\theta(t)$.}}
		\label{fig7}
	\end{minipage}
\end{figure*}
Therefore,
within compact set $\Omega_\eta(q)$, with the help of (\ref{eq58}), (\ref{eq63}), $(\ref{eq103})$ and (\ref{eq117}), $\dot V$ can be is bounded by
\begin{align}
\dot V\leq &-\sum_{i=1}^n\bar c_iz_i^2-\sum_{i=2}^n\varrho_i\hat\upsilon_i^2-\bar\delta\tilde\theta^2-\bar\sigma(\Vert \varepsilon\Vert^2+\Vert \zeta\Vert^2)\notag\\
&+\sum_{i=1}^{n-1}\frac{1}{2\phi_{i+1}}\Delta_i^2+\sum_{i=1}^nd_i+\sum_{i=1}^{n-1}\rho_{i+1}^2(\Lambda_i^2+\iota_i^2)\notag\\
\leq &
-c(\beta)V+d(\beta)\label{eq:dot_V}
\end{align}
where $c(\beta)=\min\left\{2\bar c_1,\cdots,2\bar c_n,2\varrho_2,\cdots, 2\varrho_n,2\bar \delta,\bar \sigma /\bar \lambda\right\}$ and $d(\beta)=\sum_{i=1}^{n-1}\frac{1}{2\phi_{i+1}}\Delta_i^2+\sum_{i=1}^nd_i+\sum_{i=1}^{n-1}\rho_{i+1}^2(\Lambda_i^2+\iota_i^2)$ with positive constants $\bar\sigma=1-5\sigma$ and $\bar \delta=\frac{\delta}{2}-\frac{1+\tilde\gamma_y^2}{2\sigma}$.  Note that $\bar c_1$ is defined in (\ref{eq59}), $\bar c_i=c_i-9/2\ (i=2,\cdots,n-1)$, $\bar c_n=c_n-1$, and $\bar \lambda$ is determined by $k$. Then,
there exists a $\underline q$ such that  for any given positive number $q> \underline q $, there always exists a set of parameter $\beta$ such that
$c(\beta )\geq d(\beta )/q$ (See more detailed parameter selection procedure in   Remark  \ref{rem5}).

As a result, $\dot V\leq 0$ on the boundary of  $\Omega_\eta(q)$. Due to (\ref{eq75}), $V(t)$ is continuous and no jump will occur   at $t=t_j,j=1,2,\cdots$. As a result,  $\Omega_\eta(q)$  is a  positively invariant set, i.e., if $V(t_0)\leq q$, then $V(t)\leq q$ for $t\geq t_0$.
Due to the boundedness of $z_i \ (i=1,\cdots,n)$, $\hat\upsilon_i\ (i=2,\cdots,n)$, $\tilde\theta$, $\varepsilon$ and $\zeta$, by (\ref{eq24}) and (\ref{eq123}), it follows that $\alpha_i\ (i=1,\cdots,n-1)$ and $\alpha_{if} (i=2,\cdots,n)$ are bounded and  $x$ is also bounded by (\ref{eq5}). Hence, all signals of the closed-loop systems are semiglobally bounded.

{\blue Invoking comparison lemma for \eqref{eq:dot_V} yields
\begin{align}
V(\eta(t))&\leq \exp\left(-c(\beta)(t-t_0)\right)V(\eta(t_0))\notag\\
&\quad +\frac{d(\beta)}{c(\beta)} \left(1-\exp\left(-c(\beta)(t-t_0)\right)\right)\notag
\end{align}
which implies that there exists a $\bar t$ such that  $V(\eta(t))\leq \alpha d(\beta)/c(\beta)$ for $t>\bar t$ and $\alpha>1$ and
\begin{equation}\label{eq:bound_y}
|y(t)| \leq \sqrt{{2\alpha d(\beta)}/{c(\beta)}},\quad \forall t>\bar t
\end{equation}
where we used $\frac{1}{2}y^2 \leq V_1 \leq V$. }

Finally, we show  the Zeno behavior is excluded, that is  there exist positive constants $\bar \omega$, $\omega$ such that $\bar t_{k+1}-\bar t_{k}\geq \bar \omega$, $t_{j+1}-t_j\geq \omega$ for $\forall k,j\in\mathbb{N}$. First, the derivative of $e_y$ is
\begin{equation}\label{eq124}
\frac{d e_y}{dt}= \mbox{sign}(y(t)-\bar y(t))\dot y\leq\bar\xi+q+(\bar\theta^2L_1^2+\underline\lambda\bar\theta^2)/2\triangleq \phi_y.\notag
\end{equation}
Then, $\bar \omega$ is denoted as $\bar \omega=\gamma_y/\phi_y.$
Suppose that $t_j$ is in the $J$th triggering period generated by (\ref{eq4t}), that is, $t_j\in[\bar t_{J},\bar t_{J+1})$. Note that $\bar y(t)=\bar y(\bar t_{J})=\bar y(t_j)=y(\bar t_{J})$ for $t\in[\bar t_{J},\bar t_{J+1})$, $\bar y(t)=y(\bar t_{J+1})$ for $t\in[\bar t_{J+1},\bar t_{J+2})$ and $\bar y(t)-\bar y(t_j)=\bar y(\bar t_{J+1})-\bar y(\bar t_{J})=\gamma_y$ for $t\in[\bar t_{J+1},\bar t_{J+2})$. Due to $\gamma_{\bar y}> \gamma_y$, we  obtain $\vert \bar y(t)-\bar y(t_j)\vert\leq \gamma_{\bar y}$
holds for $t\in[t_j,\bar t_{J+2})$. Due to  $\bar t_{k+1}-\bar t_{k}\geq \bar \omega$ and the fact that $t_j< \bar t_{J+1}$, we have
\begin{equation}\label{eq127}
e_{\bar y}\leq \gamma_{\bar y},\ \forall t\in[t_j,t_j+\bar \omega).
\end{equation}
The derivatives of $e_{\xi}$, $e_\zeta$, $e_{h}$, $e_{f}$ are $\dot e_{\xi}
\leq\Vert A_c\Vert \bar \xi+\Vert k\Vert \sqrt{2q}+\Vert b\Vert \bar u \triangleq \phi_{\xi}$, $\dot e_{\zeta}
\leq q+\underline\lambda\Vert A_c\Vert^2/4+\Vert L\Vert^2/2\triangleq \phi_{\zeta}$, $\dot e_h
\leq(2L_1+\underline \lambda/2+1)q+\delta^2/2+\delta\vert \theta\vert \triangleq \phi_h$, $\dot e_f\leq\Vert\left [\rho_2\upsilon_2,\cdots,\rho_n\upsilon_n\right]\Vert\leq \bar \rho\sqrt{2q}\triangleq \phi_f$.
Therefore, with (\ref{eq127}), $\omega$ can be chosen as
\begin{equation}\label{eq:zeno}
\omega=\min\left\{\bar \omega,\gamma_\xi/\phi_\xi,\gamma_\zeta/\phi_\zeta,\gamma_h/\phi_h,\gamma_f/\phi_f\right\}.\notag
\end{equation}
Thus, there is no Zeno behavior. The proof is thus  complete.
\eproof
{\blue
\brem Suppose we adopt the control scheme by using continuous value in  \eqref{eq:xt}, \eqref{eq:bar_y_and_u}, we can  change the event condition \eqref{eq94} as \begin{align}
t_{j+1}=\inf \{&t>t_j:|\kappa( \alpha_f(t_j),\hat\theta(t_j),\hat{x}(t_j),\bar y(t_j))
\notag \\
&-\kappa( \alpha_f(t),\hat\theta(t),\hat{x}(t),\bar y(t))|\geq \gamma_\kappa \}\label{eq:ED2}
\end{align}
for some positive finite constant $\gamma_\kappa$. Applying similar analysis procedures in this paper, it can be concluded that the stability result in Theorem \ref{the1} still holds.
 \erem
 }
\brem\label{rem5}
	We  will show we can find a $\underline q>0$ and  select controller parameters $\beta$ to make sure
that    $c(\beta)\geq d(\beta)/q$ for any $q>\underline q$.
  Let $\sigma< \frac{1}{5}$ be some constant such that
$\bar\sigma=1-5\sigma>0$ and $c_{\delta}>1$. Let $\eta_{0}(c_{\Delta})=2\|P\|^{2}\Vert\psi(0)\Vert^{2}/\sigma+c_{\Delta}$,  $\eta_{1}=2\bar{\sigma}/\bar{\lambda}$, $\eta_{2}(c_{\Delta})=\left(2\eta_{0}(c_{\Delta})\sigma+c_{\delta}\bar{\theta}^{2}(1+\tilde{\gamma}_{y}^{2})+\eta_{1}\bar{\theta}^{2}\sigma/2\right)/\sigma$ and
$\eta_{3}(c_{\Delta})=(c_{\delta}-1)\eta_{0}(c_{\Delta})\bar{\theta}^{2}$.  Let $\underline \eta (c_{\Delta})=\frac{\eta_{2}(c_{\Delta})+\sqrt{\eta_{2}^{2}(c_{\Delta})+4\eta_{1}(c_{\Delta})\eta_{3}(c_{\Delta})}}{2\eta_{1}(c_{\Delta})}$ and  $\underline q=\underline \eta(0)$. As a result, for any $q>\underline q$, one can find a sufficiently small $c_{\Delta}$ such that $q\geq \underline \eta (c_{\Delta})$.
 First,      we
 note that $d$ can be written as $d=\delta\bar\theta^2/2+2\Vert P\Vert^{2}\Vert\psi(0)\Vert^2/\sigma+\sum_{i=1}^{n-1}\Delta_i/(2\phi_{i+1})+\pi(\gamma)$ with  $\pi(\gamma)$ being a function increasing as each element of $\gamma$ increases. We choose sufficiently small value for each element of $\gamma$ such that $\pi(\gamma)\leq c_{\Delta}/n$. We first assume $\Delta_i/(2\phi_{i+1})\leq c_{\Delta}/n$ and postpone to show how parameter selection can guarantee it later. As a result,  $\sum_{i=1}^{n-1}\Delta_i/(2\phi_{i+1})+\pi(\gamma) \leq c_\Delta$ holds and
$d\leq \bar d:= \delta\bar\theta^2/2+ \eta_0$. Second, we  select
 $\delta= \frac{2c_\delta}{2q-\bar\theta^2}(\eta_0 + (1+\tilde\gamma_y^2)q/\sigma)$ leading to $2\bar \delta \geq c_0:= \bar d/q $.
 Note that $  \bar d/q\geq d/q$ and thus $c_0\geq d/q$.
Due to $q \geq \underline \eta(c_{\Delta})$, it leads to  $\bar \sigma/\bar \lambda \geq c_0 $.
Third, we select $\bar c_1$ such that $\bar c_1\geq c_0$ and then $c_1\geq 1+1/(4\sigma)+30\sigma L^2\tilde\gamma_y^2+5\sigma \gamma_\zeta^2+\Vert P\Vert^2L^2/\sigma+\bar c_1$. And we select $\bar c_i\geq c_0$ and then $c_i\geq 9/2+\bar c_i$ for $i=2,\cdots,n$. Last,  we show how to select $\phi$ and $\rho$ to guarantee $\Delta_i/(2\phi_{i+1})\leq c_\Delta/n$. We select
sufficiently large $\phi_2$ to make $\Delta_1/(2\phi_2)\leq c_{\Delta}/n$.
And we select $\varrho_i\geq \frac{1}{2}c_0$  and then $\rho_i\geq 2+\phi_{i}+\varrho_{i}$,   and a sufficiently large $\phi_{i+1}$ such that $\Delta_i/(2\phi_{i+1})\leq c_{\Delta}/n$  for $i=2,\cdots,n$.  As a result,    $c\geq c_0 \geq d/q$.
\erem

\section{Simulation \label{sec:sim}}
In this section, we give a simulation example to illustrate the proposed control strategy. The system is given as
$
	\dot x_1 = x_2+\theta \cos (y),\;
	\dot x_2  =  u+\theta (y+1),\; y = x_1,
$
where $\theta=1$ is an unknown parameter satisfies $\vert \theta\vert\leq \bar \theta$ with $\bar \theta=1.5$. Let $q=50$. The Lipschitz constant $L=\sqrt{2}$ and $\Vert \psi(0)\Vert=\sqrt 2$. Define trigger parameters as $\gamma=[0.05,0.051,0.2,0.2,0.2,0.2]\t $. We select $k=[5,5]\t$, then $\bar \lambda=0.1099$. $c,\varrho_2$ are selected as $c=[8.5,5.5]\t $, $\varrho_2=0.16$ according to Remark \ref{rem5}. $\phi_2$ is chosen as $\phi_2=10$ and thus we choose $\rho_2=12\geq \frac{3}{2}+\varrho_2+\phi_2$. System initial values as $x=[5,-5]\t$, $\xi=[0,0]\t$, $\zeta=[0,-4]\t$, $\hat\theta(0)=4$, $\alpha_{2f}(0)=0$, satisfying $V(\eta(0))\leq q$. The  simulation time interval is $[0,10]$. Fig. \ref{fig3} is the trajectories of $y(t)$ and $\bar{y}(t)$, where boundedness is shown. Fig. \ref{fig4} illustrates the  control input and  Fig. \ref{fig8} shows that triggering events of ED1 and ED2 are independent of each other. Fig. \ref{fig5}- Fig. \ref{fig6} show the trajectories of $\dot{\xi}(t)$, $\dot\zeta(t)$, $\dot{\theta}(t)$ and $\dot{\alpha}_{2f}(t)$. Fig. \ref{fig7} shows the boundedness of $\Vert\varepsilon\Vert$, which together with Fig. \ref{fig3}, implies the boundedness of $\xi(t)$ and $\zeta(t)$.

\begin{figure}[htbp]
	\centering
	\includegraphics[width=\linewidth]{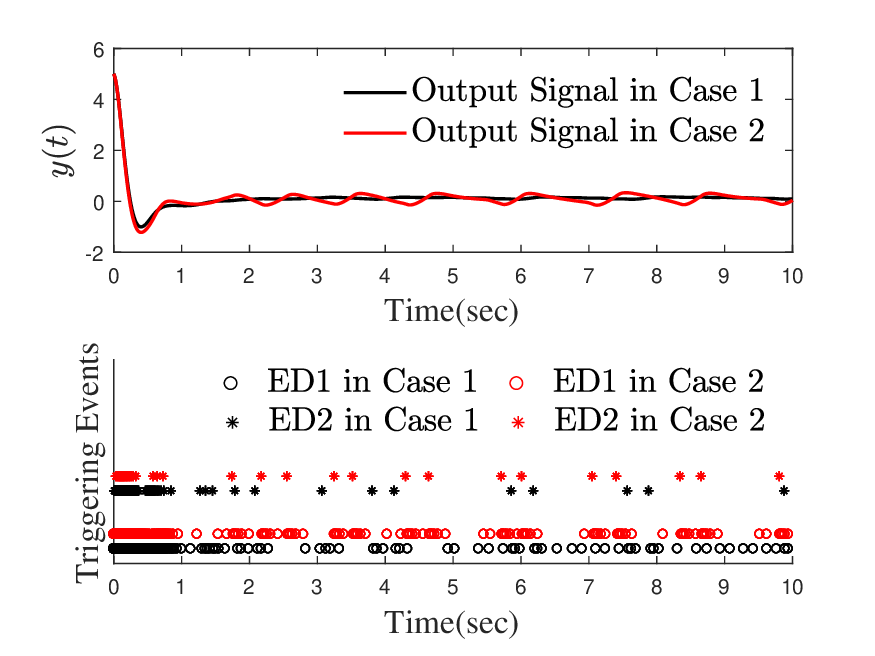}
	\caption{\small{The real-time output $y(t)$ and triggering events with different trigger conditions.}}
	\label{fig11}
\end{figure}

\vspace{2mm}
\begin{table}[htbp]
\centering
\caption{Numbers of Triggering Events}
\vspace{2mm}
\begin{tabular}{*{3}{>{\centering\arraybackslash}p{2cm}}}
\toprule[1pt]
&\normalsize{ED1} & \normalsize{ED2}\\
\midrule[0.5pt]
\normalsize{Case 1}  & \normalsize{337}& \normalsize{148}\\
\normalsize{Case 2}  & \normalsize{212}&\normalsize{37}\\
\bottomrule[1pt]
\end{tabular}
\label{table1}
\end{table}

{\blue
Now, we examine control performance under two set of triggering thresholds.
	 Case 1 is $\gamma=[0.05,0.051,0.2,0.2,0.2,0.2]\t $ and Case 2 is $\gamma=[0.3,0.31,0.5,0.5,0.5,0.5]\t $. The results are illustrated in Fig. \ref{fig11} and Table. \ref{table1}. It shows
	 different  triggering  threshold can adjust the ultimate bound of the output
	 and the number of triggering events, and there is a trade-off  between transmission frequency and control performance. The small threshold  in Case 1 increases the number of the information exchange via the network, and decreases the ultimate bound of the output.
%
\begin{figure}[htbp]
	\centering
	\includegraphics[width=\linewidth]{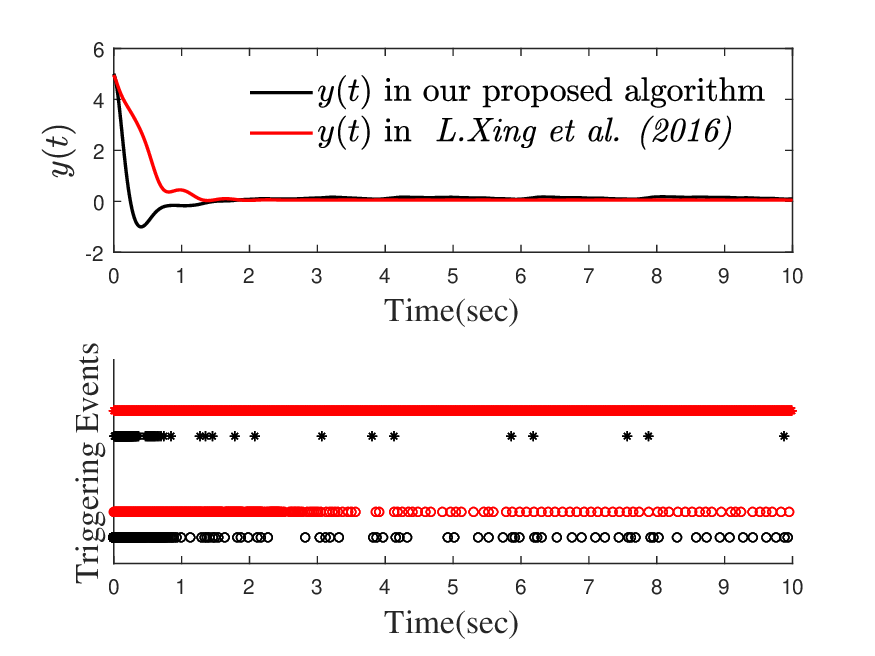}
	\caption{\small{The real-time output $y(t)$ and triggering events under our proposed algorithm and the algorithm in \cite{xing2016event04}. The black $\circ$ and $*$ denote trigger events of ED1 and ED2, respectively. The red $\circ$ and $*$  denotes the signal transmission between controller side and plant side under the algorithm in \cite{xing2016event04}. }}
	\label{fig12}
\end{figure}

\vspace{2mm}
\begin{table}[htbp]
\centering
\caption{Numbers of Triggering Events}
\vspace{2mm}
\begin{tabular}{ccc}
\toprule[1pt]
&\normalsize{Our  Algorithm} & \normalsize{Algorithm in \cite{xing2016event04} }\\
\midrule[0.5pt]
\normalsize{Controller $\Rightarrow$ Plant }  & \normalsize{337}& \normalsize{317}\\
\normalsize{Plant $\Rightarrow$ Controller}  & \normalsize{148}&\normalsize{1000}\\
\bottomrule[1pt]
\end{tabular}
\label{table2}
\end{table}
Furthermore,
we compare the proposed scheme with that in  \cite{xing2016event04} where   the full states are assumed to be accessible.
 The event-triggered controller in \cite{xing2016event04}
 is designed as
\begin{gather}
u(t)=v(t_j),\quad t\in[t_j,t_{j+1}),\quad  j\in\mathbb N\notag\\
t_{j+1} = \inf \left\{t>t_j\mid |u(t)-v(t_j)|\geq \gamma_c\right\}\notag
\end{gather}
with $\gamma_c=0.06$, $v(t)=-k z_2 -z_1 +\frac{\partial \alpha_1}{\partial x_1}x_2-\hat \theta \left(x_1 +1 -\frac{\partial \alpha_1}{\partial x_1}\right)+\frac{\partial \alpha_1}{\partial \hat \theta}\dot {\hat \theta}$, where $k=4$, $\alpha_1 =-4x_1-\hat\theta \cos(x_1)$, $z_2 =x_2 -\alpha_1$ and $\dot{\hat \theta}=x_1 \cos(x_1)+z_2\left(y+1-\frac{\partial \alpha_1}{\partial x_1}\cos(x_1)\right)-1.5\hat\theta$.
The results are shown in Fig. \ref{fig12} and Table. \ref{table2}. It is observed   that when the ultimate bound of the output is essentially the same, our proposed control algorithm requires fewer signal transmissions.
}

\section{Conclusion \label{sec:con}}
In this paper,
we propose a novel event-triggered algorithm for uncertain nonlinear systems, in which output signal, controller signal and controller dynamics are sampled to be transmitted or be updated by the two event detectors respectively at plant and controller sides. The adaptive law structure is simplified, the requirement of computation capacity at plant side is removed, and the calculation of corresponding continuous variables can be easily implemented by the simple algebraic equation. Furthermore, we solved the issue that the virtual input is no longer differentiable and proved the error bound between actual output signal and sampled transmitted output. It would be very interesting to consider the case that output at plant side is periodically sampled such that the next triggering moment at controller side can be calculated precisely.

\bibliographystyle{ieeetr}
\bibliography{ref}

\appendix
{\blue \prooflater{Lemma \ref{lem1}}
Denote the time interval $
\mathcal{H}=\mathop{\cup}\limits_{k\in \mathbb{N}}[\bar t_{k},\bar t_{k+1})$,
where $\bar t_{k}$ is the triggering instance generated by (\ref{eq4t}).
There are two cases that how the interval
$[t_j,t_{j+1}]$ is located in $\mathcal{H}$.
%
As shown in Fig. \ref{fig10}, Case 1 is  $[t_j,t_{j+1})\subseteq [\bar t_{J},\bar t_{J+1})$ for some $J$, and Case 2 is that $t_j\in [\bar t_{J},\bar t_{J+1})$ and $t_{j+1} \in [\bar t_{K},\bar t_{K+1})$ for some $\bar t_{J}$, $\bar t_{J+1}$, $\bar t_{K}$, $\bar t_{K+1}$ being triggering instances generated by (\ref{eq4t}), where $K\geq J+1$.
\begin{figure}[H]
	\centering
	\includegraphics[width=0.6\linewidth]{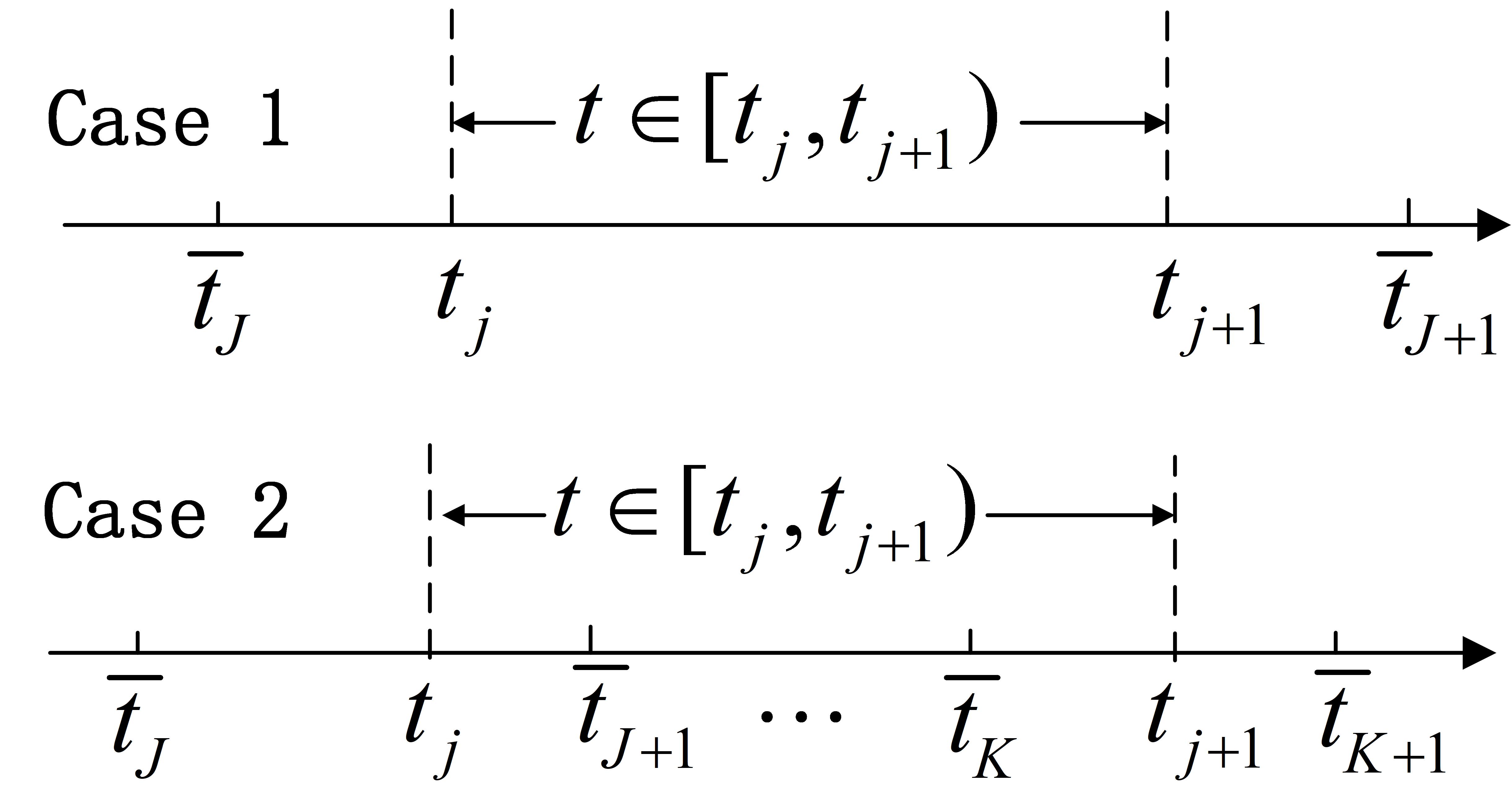}
\captionsetup{font={small}}
\vspace{0.7em}
	\caption{\small {The two cases of interval $[t_j,t_{j+1})$ in  $\mathcal{H}$.}}
	\label{fig10}
\end{figure}
In Case 1, since $\bar y(t)=y(\bar t_{J})$ holds for $t\in[\bar t_{J},\bar t_{J+1})$ and $[t_j,t_{j+1})\subseteq [\bar t_{J},\bar t_{J+1})$, $\bar y(t_j)=y(\bar  t_{J})$. By (\ref{eq4t}), one has
\begin{equation}\label{eq142}
\vert y(t)-\bar y(t_j)\vert=\vert y(t)-y(\bar t_{J})\vert\leq \gamma_y\leq \tilde\gamma_y
\end{equation}
for $t\in[t_j,t_{j+1})$.
In Case 2, when $K=J+1$, $\bar t_{J+1}=\bar t_K$ and $\bar y(t_j)=y(\bar t_J)$, one has
\begin{equation}\label{eq162}
\vert y(t)-\bar y (t_j)\vert \leq \gamma_y\leq \tilde\gamma_y, \ \forall t\in[ t_j,\bar t_{K})
\end{equation}
and
$
	\vert y(t)-\bar y(t_j)\vert= \vert y(t)-y(\bar t_{K})+y(\bar t_{K})-y(\bar t_{J})\vert
	\leq  2\gamma_y\leq \tilde\gamma_y,\ t\in[\bar t_{K},t_{j+1}).
$

When $K>J+1$, by (\ref{eq94}), $\vert \bar y(t)-\bar y(t_j)\vert\leq \gamma_{\bar y}$ holds for $t\in[t_j,t_{j+1})$.  Next, we consider the bound $|y(t)-\bar y(t_j)|$ in three intervals, namely $[t_j,\bar t_{J+1})$, $[\bar t_{J+1},\bar t_{K})$ and $[\bar t_{K},t_{j+1})$. First, from Fig. \ref{fig10}, due to
$\bar y(t_j)=y(\bar t_{J})$, one has
\begin{eqnarray}\label{eq143}
\vert y(t)-\bar y(t_j)\vert=\vert y(t)-y(\bar t_{J})\vert\leq \gamma_{ y},\ \forall t\in[t_j,\bar t_{J+1}).
\end{eqnarray}
Second,
suppose that  (\ref{eq4t}) is  triggered $l$ times in $[\bar t_{J+1},\bar t_{K})$, that is, $[\bar t_{J+1},\bar t_{K})=[\bar t_{J+1},\bar t_{J+2})\cup[\bar t_{J+2},\bar t_{J+3})\cup\cdots\cup[\bar t_{J+l},\bar t_{K})$ where $\bar t_{J+i},i=1,\cdots,l$ is triggering instances of (\ref{eq4t}). By (\ref{eq4t}), for $t\in[\bar t_{J+i},\bar t_{J+i+1})$ and $i=1,\cdots,l$ ($\bar t_K=\bar t_{J+l+1}$), there always exist a continuous function $\lambda_{i}(t)$, satisfying $\lambda_{i}(\bar t_{J+i})=0$, $\vert\lambda_{i}(t)\vert\leq 1$ for $t\in[\bar t_{J+i},\bar t_{J+i+1})$ such that $y(t)=y(\bar t_{J+i})+\lambda_{i}(t)\gamma_y$. Therefore
$
	\vert y(t)-\bar y(t_j)\vert=\vert y(\bar t_{J+i})+\lambda_{i}(t)\gamma_y-\bar y(t_j)\vert
$
holds for $t\in[\bar t_{J+i},\bar t_{J+i+1})$. Since $y(\bar t_{J+i})=\bar y(\bar t_{J+i})$ and $\bar t_{J+i}\in [t_j,t_{j+1})$, then
\begin{equation}\label{eq145}
\vert y(t)-\bar y(t_j)\vert \leq  \vert \bar y(\bar t_{J+i})-\bar y(t_j)+\lambda_{i}(t)\gamma_y\vert\leq \gamma_{\bar y}+\gamma_y
\end{equation}
holds for $t\in[\bar t_{J+i},\bar t_{J+i+1})$, and thus also holds for $t\in[\bar t_{J+1},\bar t_{K})$. Third, for $t\in[\bar t_{K},t_{j+1})$
\begin{align}\label{eq146}
\vert y(t)-\bar y(t_j)\vert&=\vert y(t)-y(\bar  t_{K})+y(\bar t_{K})-\bar y(t_j) \vert\notag\\
&\leq \gamma_{y}+\gamma_{\bar y}\leq \tilde\gamma_y.
\end{align}
By (\ref{eq142}), (\ref{eq162}), (\ref{eq143}), (\ref{eq145}) and (\ref{eq146}), we prove (\ref{eq158}).
\eproof}

\end{document}